\documentstyle[12pt]{article}
\newcommand{\bB}{{\bf B}}
\newcommand{\bC}{{\bf C}}
\newcommand{\bH}{{\bf H}}
\newcommand{\bR}{{\bf R}}

\newcommand{\cB}{{\cal B}}
\newcommand{\cD}{{\cal D}}
\newcommand{\cG}{{\cal G}}
\newcommand{\cH}{{\cal H}}
\newcommand{\cR}{{\cal R}}
\newcommand{\cQ}{{\cal Q}}
\newcommand{\cU}{{\cal U}}

\newcommand{\an}{\alpha}
\newcommand{\bn}{\beta}

\newcommand{\dn}{\delta}
\newcommand{\Psipm}{{\Psi}^{\pm)}}

\newcommand{\tpsi}{\tilde{\psi}}
\newcommand{\tPsi}{\tilde{\Psi}}
\newcommand{\tPsipm}{\tilde{\Psi}^{\pm)}}
\newcommand{\tpsipm}{\tilde{\psi}^{\pm)}}
\newcommand{\ad}{_{\alpha}}

\newcommand{\bd}{_{\beta}}

\newcommand{\aad}{_{\alpha\alpha}}
\newcommand{\bbd}{_{\beta\beta}}
\newcommand{\abd}{_{\alpha\beta}}
\newcommand{\bad}{_{\beta\alpha}}
\newcommand{\au}{^{(\alpha)}}
\newcommand{\bu}{^{(\beta)}}

\newcommand{\ju}{^{(j)}}
\newcommand{\Sum}{\sum\limits}
\newcommand{\Int}{\int\limits}
\newcommand{\Intsum}{\sum\mbox{\hskip-2em}\int\limits}
\newcommand{\intsum}[1]{\mathop{
\mbox{$\sum$\hskip-1em{\LARGE $\int$}\hskip0.1em}
}\limits_{#1}}
\newcommand{\Frac}[2]{\frac{\textstyle #1}{\textstyle #2}}
\newcommand{\Min}[1]{\mathop{{\rm min}}\limits_{#1}}
\newcommand{\Max}[1]{\mathop{{\rm max}}\limits_{#1}}
\newcommand{\Sup}[1]{\mathop{\rm sup}\limits_{#1}}
\newcommand{\opla}{\mathop{\oplus}\limits}
\newcommand{\reduction}[2]{\left. #1 \right|_{#2}}

\newcommand{\Lim}[1]{\mathop{{\rm lim}}\limits_{#1}}
\newcommand{\diag}{{\rm diag}}
\newcommand{\Img}{{\rm Im}\, }
\newcommand{\be}{\begin{equation}}
\newcommand{\ee}{\end{equation}}

\newtheorem{theorem}{\sc Theorem}
\newtheorem{lemma}{\sc Lemma}
\newtheorem{corollary}{\sc Corollary}
\newcommand{\ct}[1]{~\cite{#1}}
\newcommand{\ctt}[2]{~\cite{#1}---\cite{#2}}
\newcommand{\ctn}[1]{~\cite{#1}}
\catcode`\@=11
\def\draftlabel#1{{\@bsphack\if@filesw {\let\thepage\relax
   \xdef\@gtempa{\write\@auxout{\string
      \newlabel{#1}{{\@currentlabel}{\thepage}}}}}\@gtempa
   \if@nobreak \ifvmode\nobreak\fi\fi\fi\@esphack}
        \gdef\@eqnlabel{#1}}
\def\@eqnlabel{}
\def\@vacuum{}
\def\draftmarginnote#1{\marginpar{\raggedright\scriptsize\tt#1}}
\def\draft{\oddsidemargin -.5truein
        \def\@oddfoot{\sl preliminary draft \hfil
        \rm\thepage\hfil\sl\today\quad\militarytime}
        \let\@evenfoot\@oddfoot \overfullrule 3pt
        \let\label=\draftlabel
        \let\marginnote=\draftmarginnote
@
 \def\@eqnnum{(\theequation)\rlap{\kern\marginparsep\tt\@eqnlabel}%
\global\let\@eqnlabel\@vacuum}  }

\def\numberbysection{\@addtoreset{equation}{section}
        \def\theequation{\thesection.\arabic{equation}}}

\def\underline#1{\relax\ifmmode\@@underline#1\else
        $\@@underline{\hbox{#1}}$\relax\fi}

\catcode`@=12
\relax
\numberbysection
\begin{document}
%%%%%%%%%%%%%%%%%%%%%%%%%%%%%%%%%%%%%%%%%%%%%%%%%%%%%%%%%%%%%%%%%%%%%%
%   TITLE PAGE OF PREPRINT                                           %
%%%%%%%%%%%%%%%%%%%%%%%%%%%%%%%%%%%%%%%%%%%%%%%%%%%%%%%%%%%%%%%%%%%%%%
\baselineskip 15pt
\thispagestyle{empty}
\large
\begin{center}
{\bf Bogoliubov Laboratory of Theoretical Physics\\ }
{\bf JOINT INSTITUTE FOR NUCLEAR RESEARCH\\ }
{\bf 141980 Dubna (Moscow region), Russia}
\end{center} 
\vskip-3mm
\hrule{\hfill} 
\vskip 4.5cm
\hfill {\large Preprint JINR E5--94--259 (nucl-th/9505030)}
\bigskip

\bigskip

\bigskip

\bigskip

{\large

\noindent  A.K.Motovilov$^{1}$
}
\bigskip

\medskip

{\large

\noindent REMOVAL OF THE ENERGY DEPENDENCE \\ 
          FROM THE RESOLVENT-LIKE ENERGY-DEPENDENT \\
          INTERACTIONS$^{2}$

}
\normalsize
\bigskip

\bigskip
{\it Published in \/ 
{\rm Teoreticheskaya i Matematicheskaya Fizika~{\bf104} (1995) 
 281--303}
{\rm [}English translation in \/ {\rm Theor. Math. Phys.]}}

\vskip 5.5cm
\noindent {$^{1)}$E--mail: MOTOVILV@THSUN1.JINR.DUBNA.SU.
\newline $^{2)}$The work supported in part by the 
             International Science Foundation\newline
             (Grant~No.~RFB000).
}
\newpage
%%%%%%%%%%%%%%%%%%%%%%%%%%%%%%%%%%%%%%%%%%%%%%%%%%%%%%%%%%%%%%%%%%%%
%     ABSTRACT                                                     %
%%%%%%%%%%%%%%%%%%%%%%%%%%%%%%%%%%%%%%%%%%%%%%%%%%%%%%%%%%%%%%%%%%%%
\baselineskip15pt
\thispagestyle{empty}
\noindent {\bf Motovilov A.K.}   \hfill E5--94--259 (nucl-th/9505030)

\noindent {\bf Removal of the Energy Dependence from the Resolvent-like }

\noindent {\bf Energy-Dependent Interactions}
\bigskip

\noindent The spectral problem $(A + V(z))\psi=z\psi$ is 
considered with $A$, a self-adjoint Hamiltonian of sufficiently 
arbitrary nature.  The perturbation $V(z)$  is assumed to depend 
on the energy $z$ as resolvent of another self-adjoint operator 
$A':$ $V(z)=-B(A'-z)^{-1}B^{*}$.  It is supposed that operator 
$B$ has a finite Hilbert-Schmidt norm and spectra of operators 
$A$ and $A'$ are separated.  The conditions are formulated when 
the perturbation $V(z)$ may be replaced with an 
energy-independent ``potential'' $W$ such that the Hamiltonian 
$H=A +W$ has the same spectrum (more exactly a part of spectrum) 
and the same eigenfunctions as the initial spectral problem.  
The orthogonality and expansion theorems are proved for 
eigenfunction systems of the Hamiltonian $ H=A + W $.  
Scattering theory is developed for $H$ in the case when operator 
$A$ has continuous spectrum. Applications of the results 
obtained to few-body problems are discussed.

\medskip

The investigation has been performed at the Laboratory of Theoretical
Physics, JINR
\bigskip

\newpage
\baselineskip13pt
%%%%%%%%%%%%%%%%%%%%%%%%%%%%%%%%%%%%%%%%%%%%%%%%%%%%%%%%%%%%%%%%%%
%            THE ARTICLE ITSELF                                  %
%%%%%%%%%%%%%%%%%%%%%%%%%%%%%%%%%%%%%%%%%%%%%%%%%%%%%%%%%%%%%%%%%%
\setcounter{footnote}1
\setcounter{page}1
%%%%%%%%%%%%%%%%%%%%%%%%%%%%%%%%%%%%%%%%%%%%%%%%%%%%%%%%%%%%%%%%%%%%%%
\section{\hspace*{-1em}. INTRODUCTION}\label{Intro}
%%%%%%%%%%%%%%%%%%%%%%%%%%%%%%%%%%%%%%%%%%%%%%%%%%%%%%%%%%%%%%
Perturbations, depending  on   the   spectral
parameter  (usually energy of system)  arise
in  a  lot   of   physical
problems (see papers\ctt{Dashen}{IMU} and Refs. therein).
In  particular,  such  are  the  interaction
potentials between clusters formed by quantum
particles\ctt{Dashen}{NarodetskyPalanga}.

The  perturbations  of this type appear
typically\ctt{Dashen}{Kerbikov},\ctt{Pavlov84}{IMU}
as a result  of
dividing the Hilbert space $\cH$ of physical system in two subspaces,
$\cH=\cH_1\oplus\cH_2$.
The first one, say $\cH_1$, is interpreted as a space of
``external'' (for example, hadronic) degrees of freedom. The second one,
$\cH_2$, is associated with an ``internal'' (for example, quark) structure
of the system. The Hamiltonian ${\bH}$ of the system looks as a matrix,
\be
\label{twochannel}
{\bH}=\left[
\begin{array}{cc}
                     A_1              &          B_{12}          \\
                     B_{21}           &          A_2
\end{array}
\right]
\ee
with $A_1$, $A_2$, the channel Hamiltonians (self-adjoint
operators) and
$B_{12}$, $B_{21}=B_{12}^{*}$, the coupling operators.
Reducing the spectral problem
${\bH}U=zU$ , $U=\{ u_1 ,u_2 \}$ to the
channel $\an$ only one gets the spectral problem
\be
\label{ini}
[A\ad  +V\ad  (z)]u\ad  =zu\ad  ,
\quad \an =1,2,
\ee
where the perturbation
\be
\label{epot}
V\ad  (z)=-B\abd (A\bd -z)^{-1}B\bad ,
\quad \bn \neq \an,
\ee
depends on the spectral parameter $z$ as the resolvent $(A\bd - z)^{-1}$ of
the Hamiltonian $A\bd$.
In more complicated cases $V\ad(z)$ can include also linear terms in respect
with  $z$. Other types of dependency of the potentials
$V\ad(z)$ on the spectral parameter z give, in a general way,
the spectral problems~(\ref{ini}) with a complex spectrum.

The present paper is a continuation of the author's
works\ctt{MotJMPh91}{SPbWorkshop} devoted to a study of the
possibility to ``remove'' the energy dependence from  perturbations
of the type~(\ref{epot}). Namely, in\ctt{MotJMPh91}{SPbWorkshop}
we construct such new potentials $W\ad$ that spectrum of the Hamiltonian
$H\ad=A\ad+W\ad$ is a part of the spectrum of the problem~(\ref{ini}).
At the same time, the respective eigenvectors of $H\ad$ become
also those for~(\ref{ini}). Hamiltonians $H\ad$ are found as solutions of the
non-linear operator equations
\be
\label{basic}
H\ad=A\ad + V\ad (H\ad)
\ee
first appeared in the paper\ct{Braun} by  M.A.Braun in connection with consideration
of the quasipotential equation. The operator-value function
$V\ad(Y)$ of the operator variable $Y$, $Y:\, \cH\ad\rightarrow\cH\ad,$ is defined
by us in such a way (see Sec.~\ref{BasicEquation}) that eigenvectors $\psi$ of $Y$,
$Y\psi =z\psi$, become automatically those for $V\ad(Y)$ and
$V\ad(Y)\psi=V\ad(z)\psi.$

In Ref.~\ctn{MotJMPh91}, the case is considered in details when one of the
operators $A\ad$ is the Schr\"{o}dinger operator in $L_2(\bR^n)$ and another
one has a discrete spectrum only. The
reports\ct{Kazimierz},\ct{SPbWorkshop}  announce the results concerning
the equations~(\ref{basic}) and properties of their solutions $H\ad$ in
a rather more general situation when the Hamiltonian ${\bH}$ may be rewritten
in terms of a two-channel variant of the Friedrichs model
investigated by O.A.Ladyzhenskaya and L.D.Faddeev in
Refs.~\ct{LadyzhFaddeev},\ct{Faddeev64}. In Ref.~\ct{IMU}
the method\ctt{MotJMPh91}{SPbWorkshop} is used to construct an
effective cluster Hamiltonian for atoms adsorbed by the metal surface.

In the present paper, we specify the assertions
from\ct{Kazimierz},\ct{SPbWorkshop}
and give proofs for them.
Also, we pay attention to an important circumstance
disclosing a nature of solutions of the basic equations
(\ref{basic}). Thing is that
the potentials $W\ad=V\ad(H\ad)$ may be presented
in the form $W\ad=B\abd Q\bad$ where the operators $Q\bad$ satisfy the
equations~(\ref{QbasicSym}) (see Sec.~\ref{BasicEquation}). Exactly the same equations arise
in the method of construction of invariant subspaces for self-adjoint
operators developed by V.A.Malyshev and R.A.Minlos in
Refs.~\ct{MalyshevMinlos1},\ct{MalyshevMinlos2}.
It follows from the results of\ct{MalyshevMinlos1},\ct{MalyshevMinlos2}
that operators $H\ad$, $\an=1,2,$ determine in fact,  parts of the
two-channel Hamiltonian ${\bH}$ acting in corresponding invariant subspaces
(see Theorem~\ref{ThInvariant} and comments to it).

Recently, the author came to know about the work\footnote{Accepted 
for publication in J.~Operator Theory.}
``Spectral properties of a class
of rational operator-value functions''
by  V.M.Adamyan and H.Langer studying  
the operator-value functions written in our notation as 
$F\ad(z)=z-A\ad\pm B\abd(A\bd-z)^{-1}B\bad$. In particular 
Adamyan and Langer show in this work 
that a subset of eigenvectors of $F\ad$ can be chosen 
to form a Riesz basis in 
$\cH\ad$.  There is a certain intersection of their results and ours from 
Refs.~\ctt{MotJMPh91}{SPbWorkshop}. However the methods are different.

The paper is organized as follows.

In Sec.~\ref{Initial} we describe the Hamiltonian ${\bH}$ as a
two-channel variant of the Friedrichs
model\ct{LadyzhFaddeev}, \ct{Faddeev64}.
We suppose that both operators $A\ad$, $\an=1,2,$ may have continuous
spectrum. When properties of objects connected with this spectrum
(wave operators and scattering matrices) are considered in
following sections, the coupling operators $B\abd$ in~(\ref{twochannel})
are assumed to be integral ones with kernels $B\abd(\lambda,\mu)$, the
 H\"{o}lder functions in both variables $\lambda,\mu$.

In Sec.~\ref{BasicEquation} the equations~(\ref{basic}) are studied. As in
Refs.~\ct{MalyshevMinlos1},\ct{MalyshevMinlos2} we suppose that
spectra $\sigma(A_1)$ and $\sigma(A_2)$ of the operators $A_1$ and
$A_2$ are separated, ${\rm dist}\{ \sigma(A_1), \sigma(A_2) \}>0$.
Existence of solutions of Eqs.~(\ref{basic}) is established
only in the case when the Hilbert-Schmidt norm $\| B\abd\|_2$
of the coupling operators satisfies the condition\newline
$\| B\abd\|_2 < \Frac{1}{2}{\rm dist}\{ \sigma(A_1),\sigma(A_2) \} $.

In Sec.~\ref{Eigenfunctions} the eigenfunctions systems
of the operators $H\ad$ are studied and  theorems of their
orthogonality and completeness are proved. We show here
in particular that spectrum of the Hamiltonian ${\bH}$ is distributed
between the solutions $H_1=A_1 + B_{12} Q_{21}$ and
$H_2=A_2 + B_{21} Q_{12}$, $Q_{21}=-Q_{12}^{*}, $
of the basic equations~(\ref{basic})
in such a way that $H_1$ and $H_2$ have not ``common'' eigenfunctions
$U=\{u_1, u_2\} $ of ${\bH}$: simultaneously, component $u_1$ can not be
eigenfunction for $H_1$, and component $u_2$,  for $H_2$.

In Sec.~\ref{InnerProduct} we introduce new inner products
in the Hilbert spaces $\cH\ad$, $\an=1,2,$
making the Hamiltonians $H\ad$ self-adjoint.

In Sec.~\ref{ScatteringProblem}
we give a non-stationary formulation of the scattering
problem for a system described by the Hamiltonians $H\ad$ constructed.
We show that this formulation is correct and scattering operator is
exactly the same as in initial spectral problem.

At last, in Sec.~\ref{2bodyPotNbody}
we discuss the questions concerning a use of two-body
energy-dependent potentials in few-body problems.
%%%%%%%%%%%%%%%%%%%%%%%%%%%%%%%%%%%%%%%%%%%%%%%%%%%%%%%%%%%%%%%%%%%
\section{\hspace*{-1em}. INITIAL SPECTRAL PROBLEM AND \newline 
                         TWO-CHANNEL HAMILTONIAN}
\label{Initial}
%%%%%%%%%%%%%%%%%%%%%%%%%%%%%%%%%%%%%%%%%%%%%%%%%%%%%%%%%%%%%%%%%%%
Let $A_{1}$ and $A_{2}$ be self-adjoint  operators  acting,
respectively, in  ``external'', ${\cH} _{1}$,  and  ``internal'',
${\cH} _{2}$, Hilbert spaces. We study the spectral problem~(\ref{ini})
with  perturbation $V\ad(z)$  given by~(\ref{epot}).
We suppose that $B\abd \in {\bB} ({\cH} \ad  ,{\cH} \bd )$ where
${\bB} ({\cH} \ad  ,{\cH} \bd )$ is the
Banach  space  of   bounded  linear
operators acting from ${\cH} \ad  $ to ${\cH} \bd $.

Note that method developed in the present paper works also in the
case of more general perturbations\footnote{Remember
%%%%%%%%%%%%%%%%%%%%%%%%%%%%%%%%%%%%%%%%%%%%%%%%%%%%%%%%%%%%%%%%%%%%%
        that if $N\ad\geq 0$
        then Eq.~(\ref{Rfunction}) gives a
        general form of $R$-function on $\cH\ad$, i.e. an analytic at
        $\Img z\neq 0$  ${\bB} (\cH\ad,\cH\ad)$--value function with
       positive imaginary part for $z:\, \Img z > 0$
       (see paper\ct{Naboko} and Refs. therein).   }
%%%%%%%%%%%%%%%%%%%%%%%%%%%%%%%%%%%%%%%%%%%%%%%%%%%%%%%%%%%%%%%%%%%%%%
 $V\ad(z)=-\cR\ad(z)$ containing linear terms,
\be
\label{Rfunction}
  \cR\ad(z)= N\ad z + B\abd(A\bd-N\bd z -z)^{-1}B\bad
\ee
with  $N\ad$, self-adjoint bounded operator
in $\cH\ad$ such that $N\ad\geq (\dn -1)I\ad$ where
$\,\,\dn  > 0 $ and $I\ad$ is the identity operator in $\cH\ad$.
Thing is that the equation~(\ref{ini}) with
$V\ad(z)=-\cR\ad(z)$ can be easily rewritten in the form
(\ref{ini}),(\ref{epot}). To do this, one has only to make the replacements
$u\ad\rightarrow u\ad'=(I\ad+N\ad)^{1/2}u\ad,$
$
\,\, A\ad\rightarrow A\ad'=(I\ad+N\ad)^{-1/2} A\ad (I\ad+N\ad)^{-1/2}
$
and
$
        B\abd\rightarrow B\abd'=
  (I\ad+N\ad)^{-1/2}B\abd (I\bd+N\bd)^{-1/2}
$.
Therefore we shall consider further only the initial spectral problem
(\ref{ini}),(\ref{epot}).

We shall assume that operators $A\ad$, $\an=1,2,$ may have continuous
spectra $\sigma\ad^c$. To deal with these spectra we accept below some
presuppositions with respect to $A\ad$ restricting us to the case of
a two-channel variant of the Friedrichs
model\ct{LadyzhFaddeev}, \ct{Faddeev64}.
Note that these presuppositions are not necessary
for a part of statements (Lemma~\ref{LQdefinition},
Theorems~\ref{ThSolvability} --- \ref{ThDiscSpectrum}
and~\ref{ThInnerProduct}) which stay
correct also in general case.

The presuppositions are following.

At first, we assume that
Hamiltonian $H$ is defined  in  that  representation  where
operators $A\ad  , \an =1,2,$ are diagonal. We suppose
that continuous  spectra $\sigma ^{c}\ad  $
of  the  operators $A\ad  ,
\an =1,2,$ are absolutely continuous and consist of a finite
number of finite (and may be one or two infinite) intervals
$(a\ad\ju ,b\ad\ju )$,
$ -\infty\leq a\ad\ju  < b\ad\ju  \leq +\infty$,
$j=1,2,\ldots,n\ad,$ $n\ad < \infty$.
At second, we suppose that discrete spectra $\sigma\ad^d$ of
the operators $A\ad$, $\an=1,2,$ do not intersect with $\sigma\ad^c$,
$\sigma\ad^d \bigcap\sigma\ad^c =\emptyset$,
and consist of a finite number of points with finite multiplicity.
In  this  case  the
space ${\cH} \ad  $  may be present as the direct
integral\ct{BirmanSolomiak}
\be
\label{Neumann}
{\cH} \ad
=\Intsum_{\lambda\in\sigma\ad }
\oplus {\cG} \ad  (\lambda )d\lambda
\equiv \opla_{\lambda\in\sigma^{d}\ad  }
{\cG} \ad  (\lambda )\oplus
\Int_{\lambda\in\sigma\ad^c}
\oplus{\cG} \ad  (\lambda )d\lambda , \quad
\sigma\ad =\sigma^{c}\ad\bigcup\sigma^{d}\ad\subset\bR .
\ee
The space ${\cH} \ad  $
consists of the
measurable functions $f\ad  $
which are defined  on $\sigma \ad  $
and
have the values
$f\ad  (\lambda )$
from  corresponding  Hilbert
spaces               ${\cG} \ad  (\lambda )$.
By $\langle\,\cdot\, ,\,\cdot\,\rangle$
we denote the inner  product  in
${\cH} \ad  ,\qquad $
$$
\langle f\ad  ,g\ad  \rangle=
\intsum{\lambda\in\sigma\ad  }
(f\ad  (\lambda ),g\ad  (\lambda ))
\equiv \Sum_{\lambda\in\sigma^{d}\ad  }
(f\ad  (\lambda ),g\ad  (\lambda ))
+\Int_{\lambda\in\sigma\ad^c}
d\lambda (f\ad  (\lambda ),g\ad  (\lambda )),
$$
where $(\,\cdot\, ,\,\cdot\,)$ stands for inner product in
${\cG} \ad  (\lambda )$. By  $|\cdot| $   we  denote
  norm  of  vectors  and
operators in ${\cG} \ad  (\lambda )$ and by
$\parallel\cdot\parallel $,
the  norm  in ${\cH} \ad  $.
 Operator $A\ad  $
acts in ${\cH} \ad  $
as  the  independent  variable multiplication operator,
\be
\label{multi}
(A\ad  f\ad  )(\lambda )=
\lambda \cdot f\ad  (\lambda ), \quad \an =1,2.
\ee
It's domain
${\cal D}(A\ad  )$
consists of  those  functions $f\ad  \in {\cH} \ad  $
which satisfy the condition
\newline
$\intsum{\lambda\in\sigma \ad  }\lambda ^{2}$
$|f\ad  (\lambda )|^{2}<\infty $. For  the
sake of simplicity we assume that
${\cG} \ad(\lambda )$ does not depend on
$\lambda\in\sigma\ad^c$, i.e.
$
{\cG} \ad(\lambda )\equiv {\cG} ^{c}\ad
$
for each
$
\lambda \in \sigma^{c}\ad  .
$
Hence,
$
\Int_{\sigma ^{c}\ad  }
\oplus {\cG} \ad  (\lambda )d\lambda =
L_{2}(\sigma ^{c}\ad  ,{\cG} ^{c}\ad  )
\equiv {\cH} ^{c}\ad
$.
By $E\ad  (d\lambda )$  we  denote  a
spectral   measure\ct{BirmanSolomiak}   of
the operator $A\ad  ,$
$
       A\ad  =\Int_{\sigma\ad}\lambda E\ad  (d\lambda )
$.
In the diagonal representation  considered,
the spectral projector $E\ad  $ acts on $f\in {\cH} \ad  $ as
\be
\label{mera}
(E\ad  (\Delta  )f)(\lambda )=\chi_{\Delta  }(\lambda )f(\lambda )
\ee
for any Borelian set
$
\Delta  \subset \sigma \ad
$.
Here, $\chi _{\Delta  }$
is a characteristic function of
$\Delta  ,$
$ \chi _{\Delta  }(\lambda )=1$
if
$\lambda \in \Delta  $,
and $\chi _{\Delta  }(\lambda )=0$
if
$\lambda \not\in \Delta  $.

Let ${\cB} ^{\an\bn}_{\theta\gamma}$
be a class of functions $F$  defined  on
$\sigma \ad  \times \sigma \bd ,$ $\an,\bn=1,2$,
for    each $\lambda \in \sigma \ad  , \mu \in \sigma \bd $
as   operator
$F(\lambda ,\mu ):$
${\cG} \bd (\mu )\rightarrow {\cG} \ad  (\lambda )$,
with $\parallel F\parallel_{{\cB} } < \infty $,  where
$$
\parallel F\parallel _{{\cB} } =
\Sup{
\mbox{
\scriptsize
$\begin{array}{c}
\mu\in\sigma\bd\\
\lambda\in\sigma\ad
\end{array}$
}
}(1+|\lambda |)^{\theta }
(1+|\mu |)^{\theta }| F(\lambda ,\mu ) |  +
$$
$$+\Sup{
\mbox{
\scriptsize
$\begin{array}{c}
\lambda,\lambda'\in\sigma\ad^c\\
\mu\in\sigma\bd
\end{array}$
}
}
\left\{
(1+|\mu |)^{\theta}
\Frac{|F(\lambda ,\mu )-F(\lambda ',\mu )|}{|\lambda  -\lambda '|^{\gamma }}
\right\}+
$$
$$
+\Sup{
\mbox{
\scriptsize
$\begin{array}{c}
\lambda\in\sigma\ad\\
\mu,\mu'\in\sigma\bd^c
\end{array}$
}
}
\left\{   (1+|\lambda|)^{\theta }
\Frac{|F(\lambda ,\mu )-F(\lambda ,\mu ')|}{|\mu-\mu'|^{\gamma }}
\right\}+
$$
$$
+\Sup{
\mbox{
\scriptsize
$\begin{array}{c}
\lambda,\lambda'\in\sigma\ad^c\\
\mu,\mu'\in\sigma\bd^c
\end{array}$
} }
\left\{ \Frac{|F(\lambda,\mu)-F(\lambda',\mu)-
F(\lambda,\mu')+F(\lambda',\mu')|}
{|\lambda-\lambda'|^{\gamma}|\mu-\mu'|^{\gamma }}
\right\}.
$$
With the norm $\parallel\cdot\parallel_{{\cB} }$
this class will constitute a Banach space.
We introduce also the Banach space
${\cal M}_{\theta \gamma }(\sigma \ad  )$  of
functions $f$ defined on $\sigma \ad  $ with the norm
$$
\parallel f\parallel_{{\cal M}}
=\Sup{\lambda\in\sigma\ad}(1+|\lambda |)^{\theta }|f(\lambda)|
+\Sup{\lambda,\lambda'\in\sigma\ad^c}
\Frac{|f(\lambda ) - f(\lambda ')|}
{|\lambda  -\lambda '|^{\gamma }}<\infty .
$$
 The  value $f(\lambda )$
of  the  function $f\in {\cal M}_{\theta \gamma }(\sigma \ad  )$
is  an operator in ${\cG} \ad  (\lambda )$.

Let $B\abd $ be an integral  operator  with a
kernel $B\abd (\lambda ,\mu )$
from the space ${\cB} ^{\an \bn }_{\theta \gamma },$
$\theta >{1\over 2}, {1\over 2}<\gamma <1.$  We assume that
$B\abd (\lambda ,\mu )$ is
 a compact operator, $B\abd (\lambda ,\mu ):$
${\cG} \bd (\mu )\rightarrow {\cG} \ad  (\lambda )$,
for each $\lambda \in \sigma \ad  , \mu \in \sigma \bd $ and
 $B\abd (\lambda ,\mu )=0$
if $\lambda $ belongs to the boundary of $\sigma ^{c}\ad  $
or $\mu $ belongs to the
boundary of $\sigma ^{c}\bd $.

With this presuppositions the Hamiltonian ${\bH}$
may be considered as a two-channel variant of the Friedrichs
model\ct{LadyzhFaddeev},\ct{Faddeev64}. Investigation of ${\bH}$
repeats almost literally the analysis from
Ref.~\ct{Faddeev64}. Therefore we describe
here only final results which are quite analogous to
\ct{LadyzhFaddeev},\ct{Faddeev64}. These results are
following.

The  operator ${\bH}$  is  self-adjoint  on  the  set
${\cal D}({\bH})={\cal D}(A_{1})\oplus {\cal D}(A_{2})$.
Continuous  spectrum  of ${\bH}$  is
situated on the set $\sigma _{c}({\bH})=\sigma ^{c}_{1}\cup \sigma ^{c}_{2}$.
Let ${\bH}^{c}$ be the part of ${\bH}$
acting in the invariant  subspace  corresponding  to
continuous  spectrum.   The  operator ${\bH}^{c}$
is  unitary equivalent  to   the   operator
${\bH}_{0}=A^{(0)}_{1}\oplus A^{(0)}_{2}$   with
$A^{(0)}\ad,$ $\an=1,2,$ the restriction of the operator
$A\ad$ on $\cH^{c}\ad$.
Namely,  there   exist  wave  operators $U^{(+)}$   and $U^{(-)}, $
$
U^{(\pm )}=
\left(
\begin{array}{lr}
u^{(\pm )}_{11}    &     u^{(\pm )}_{12}  \\
u^{(\pm )}_{21}    &     u^{(\pm )}_{22}
\end{array}
\right)=
$
$
s\!-\!\Lim{t\rightarrow\mp\infty}{\rm e}^{i{\bH}t}{\rm e}^{-i{\bH}_0 t},
$
 with  the   following   properties:
${\bH}U^{(\pm )}=U^{(\pm )}{\bH}_{0},$ $U^{(\pm )*}U^{(\pm )}=I,$
$U^{(\pm )}U^{(\pm )*}=  I-P.$
Here, $P$ is an
orthogonal projector on  subspace
corresponding to the discrete spectrum $\sigma _{d}({\bH})$
of the operator ${\bH}$.

The kernel $u^{(\pm )}_{\an \an }(\lambda ,\lambda ')$
of the operator $u^{(\pm )}_{\an \an },\an =1,2,$
represents an eigenfunction of the continuous spectrum
of the problem~(\ref{ini}) for
$
z=\lambda '\pm i0,
$
$
\lambda '\in \sigma ^{c}\ad
$, and  satisfies the integral equation
\be
\label{wavefunction}
u^{(\pm )}_{\an \an }(\lambda ,\lambda ')=
I\ad^c  \dn  (\lambda -\lambda ')-
[(A\ad  -\lambda '\mp i0)^{-1}
V\ad  (\lambda '\pm i0)u^{(\pm )}_{\an \an }]
(\lambda ,\lambda '),
\ee
where $I\ad^c  $
is identity operator  in
$
{\cG} ^{c}\ad  , \lambda \in \sigma \ad
$.
For each concrete  sign  (plus  or  minus)  and  for  each
$
\lambda '\in \sigma ^{c}\ad  , \lambda '\not\in \sigma _{d}({\bH})
$
the function
$
u^{(\pm )}_{\an \an }(\lambda ,\lambda ')
$
is  an  unique solution of eq.(\ref{wavefunction})
in the class of the distributions
\be
\label{amplitude}
f^{(\pm )}\ad  (\lambda )=
I\ad  \dn  (\lambda -\lambda ')+
{f(\lambda )\over \lambda -\lambda '\mp i0},
\quad f\in {\cal M}_{\theta '\gamma '},
\ee
where
$
{1\over 2}<\theta '<\theta ,
$
$
{1\over 2}<\gamma '<\gamma
$.
At the same time
$$
u^{(\pm )}\abd (\lambda ,\lambda ')=
-[(A\ad  -\lambda '\mp i0)^{-1}
B\abd u^{(\pm )}_{\bn \bn }](\lambda ,\lambda '),
\quad \bn \neq \an ,
$$
is the  problem ~(\ref{ini})  eigenfunction  corresponding  to
$\lambda '\in \sigma\bd^c$.

The functions $u_{\bn\an}^{(\pm)}$, $\an,\bn=1,2,$ can be explicitly
expressed in terms of kernels of the operator
$$
T(z)=B-B({\bH}-z)^{-1}B,\quad
B=\left[\begin{array}{lr}   0       &   B_{12}  \\
                           B_{21}   &     0     \end{array}\right].
$$
Corresponding formulae read as
$$
u_{\bn\an}^{(\pm)}(\lambda,\lambda')=
\dn _{\bn\an}I\ad^c\dn (\lambda-\lambda')-
\Frac{T_{\bn\an}(\mu,\lambda',\lambda'\pm i0)}
{\mu-\lambda'\mp i0}, \quad
\mu\in\sigma\bd,\,\, \lambda'\in\sigma\ad^c,
$$
with $t$-matrices
$$
T_{\an\an}=B\abd\left[z-A\bd+B_{\bn\an}(A\ad-z)^{-1}B\abd\right]^{-1}
B_{\bn\an}
$$
      and
$$
T_{\bn\an}=B_{\bn\an}\left[z-A\ad+B_{\an\bn}(A\bd-z)^{-1}
B_{\bn\an}\right]^{-1}(z-A\ad)=
$$
$$
=(z-A\bd)\left[z-A\bd+B_{\bn\an}(A\ad-z)^{-1}B\abd\right]^{-1}
B_{\bn\an},\quad \bn\neq\an.
$$
Considering the equation for $T(z)$,
$T(z)=B-B(A-z)^{-1}T(z)$, $A=A_1\oplus A_2$, one shows
in the same way as in\ct{LadyzhFaddeev},\ct{Faddeev64}
that for all $z\in\bC\setminus\sigma({\bH})$, each kernel
$ T_{\bn\an}(\mu,\lambda,z)$, $\an,\bn=1,2,$
belongs to the class
${\cB} _{\theta'\gamma'}^{\bn\an}$
with arbitrary $\theta',\gamma'$
such that $\Frac{1}{2}<\theta'<\theta$,
$\Frac{1}{2}<\gamma'<\gamma$.
In respect with variable $z$,
the kernel of $T_{\bn\an}(z)$
is continuous in the ${\cB} _{\theta'\gamma'}^{\bn\an}$--norm
right up to the upper and lower borders of the set
$\sigma_c({\bH})\setminus\sigma_d({\bH})$.

Scattering operator $S=U^{(-)*}U^{(+)}$ for a system described by the
Hamiltonian ${\bH}$ is unitary in $\cH\ad^c$. It's kernels
$s_{\bn\an}(\mu,\lambda)$, $\an,\bn=1,2$, are given by expressions
\be
\label{scattering}
  s_{\bn\an}(\mu,\lambda)=\dn (\mu-\lambda)
  \left[       \dn _{\bn\an}I\ad^c-
  2\pi i\, T_{\bn\an}(\mu,\lambda,\lambda+i0)  \right].
\ee

By $U_{j},$ $j=1,2,\ldots$,
we  denote  eigenvectors,
$U_{j}=\{u\ju _{1},u\ju _{2}\},$
$ U_{j}\in {\cal D}({\bH})$,
$\parallel U_{j} \parallel =1,$ and by $z_{j},$ $ z_{j}\in {\bf R}$, the
respective eigenvalues of  the   operator ${\bH}$  discrete
spectrum $\sigma_d({\bH})$. The component
$u\ju \ad  ,$ $\an =1,2,$ of the  vector $U_{j}$
is a solution of  Eq.~(\ref{ini})  at $z=z_{j}$.
If $z_{j}\!\in\!\sigma ^{c}\bd $  then
$(B\bad u\ju \ad  )(z_{j})=0.$
%
%%%%%%%%%%%%%%%%%%%%%%%%%%%%%%%%%%%%%%%%%%%%%%%%%%%%%%%%%%%%%%%%%%%%%%%%%
\section{\hspace*{-1em}. BASIC EQUATION}\label{BasicEquation}
%%%%%%%%%%%%%%%%%%%%%%%%%%%%%%%%%%%%%%%%%%%%%%%%%%%%%%%%%%%%%%%%%%%%%%%%%
%
The paper is  devoted  to  construction  of  such
operator $H\ad  $
that it's each eigenfunction
$u\ad  , H\ad  u\ad  =zu\ad  $,
together with eigenvalue $z$,  satisfies  Eq.~(\ref{ini}).  This
operator will be found as a solution of the non-linear
operator equation~(\ref{basic}). To obtain this equation we need
the following operator-value function $V\ad  (Y)$
of the  operator
variable $Y:\quad$
%%%%%
$$
V\ad  (Y)=B\abd\Int_{\sigma \bd }
E\bd (d\mu )B\bad (Y-\mu )^{-1},
$$
%%%%%
$Y:$ ${\cH} \ad  \rightarrow {\cH} \ad  $ .
We suppose here that
%%%%%
$
(Y-\mu I)^{-1}\in L_{\infty }
(\sigma \bd ,{\bB} ({\cH} \ad  ,{\cH} \ad  ))
$
if $\mu \in \sigma \bd $.
This means that $\sigma\bd$ has not to be included into the spectrum of
the operator $Y$. Integral
$
Q(T)=\Int_{\sigma \bd }
E\bd (d\mu )B\bad T(\mu )
$
for
$
T\in L_{\infty }(\sigma \bd ,{\bB} ({\cH} \ad  ,
{\cH} \ad  )),
$
$
\|{T}\|_{\infty }=
$
$
E\bd\!-\!\Sup{\mu\in\sigma\bd}\|T(\mu)\|<\infty
$,
is  constructed  in the same way as integrals of scalar functions
 over spectral measure (see
Ref.\ct{BirmanSolomiak}, p.130). Namely as a limit value,
in respect to the operator
norm in ${\bB} ({\cH} \ad  ,{\cH} \ad  )$,
of  respective finite integral
sums for   piecewise-constant  operator-value  functions
approximating $T$  in
$
L_{\infty }(\sigma \bd ,{\bB}
({\cH} \ad  ,{\cH} \ad  ))
$.
We  show   the
existence of this  integral at least in the case when the
Hilbert-Schmidt  norm
$\|{B}_{\an\bn}\|_{2}$ is finite.

%%%%%%%%%%%%%%%%%%%%%%%%%%%%%%%%%%%%%%%%%%%%%%%%%%%%%%%%%%%%%%%%%%%%%%
\begin{lemma}\label{LQdefinition}\hspace*{-0.5em}{\bf :}
%%%%%%%%%%%%%%%%%%%%%%%%%%%%%%%%%%%%%%%%%%%%%%%%%%%%%%%%%%%%%%%%%%%%%%
 Let
$
T\in L_{\infty }(\sigma \bd ,
{\bB} ({\cH} \ad  ,{\cH} \ad  ))
$
 and
$
\parallel B\abd \parallel_{2}<\infty $.
 Then  the
integral $Q(T)$   exists  being  a   bounded   operator,
$Q(T):$ ${\cH} \ad  \rightarrow {\cH} \bd $,
$
\parallel Q(T)\parallel
$
$
\le \parallel T\parallel _{\infty }\!\cdot\!
 \parallel B\bad \parallel_{2}
$.
\end{lemma}

\noindent{\sc Proof}. We prove the  Lemma  in  the
diagonal  representation~(\ref{Neumann}),~(\ref{multi}).
By~(\ref{mera}) we have
$$
(Qf)(\mu )=
\intsum{\sigma\ad} B\bad (\mu ,\lambda )
(T(\mu )f)(\lambda) d\lambda
$$
for
any $f\in {\cH} \ad  $.
It means that
$$
| (Qf)(\mu )|^{2}\!\le\!\Intsum_{\lambda\in\sigma\ad}
d\lambda \cdot
| B\bad (\mu,\lambda )|^{2}
\Intsum_{\lambda\in\sigma\ad}d\lambda |({T(\mu)}f)(\lambda)|^{2}=
$$
$$
=\Intsum_{\lambda\in\sigma\ad} d\lambda|B_{\bn\an}(\mu,\lambda)|^{2}\cdot
\|T(\mu)f\|^2
\leq \Intsum_{\lambda\in\sigma\ad}d\lambda|B_{\bn\an}(\mu,\lambda)|^{2}\cdot
\|T(\mu)\|^{2}\cdot\|f\|^2.
$$
Hence,
integrating  over
$\mu \in \sigma\bd $
we  come  to  the   relation
$$
\parallel{Qf}\parallel^{2}\le\parallel{B_{\bn\an}}\parallel^2_2\!\cdot\!
\parallel{T}\parallel^{2}_{\infty }\!\cdot\!
\parallel{f}\parallel^{2}
$$
which completes the proof.
\medskip

Let us suppose that
$
(H\ad  -\mu I)^{-1}\in L_{\infty }
(\sigma \bd ,{\bB} ({\cH} \ad  ,{\cH} \ad  ))
$.
We note that if
$H\ad  \psi\ad  =z\psi\ad  $,
then automatically
$$
V\ad  (H\ad  )\psi\ad  =
B\abd\Int_{\sigma \bd }
E\bd (d\mu )
B\bad (z-\mu )^{-1}\psi\ad  =
$$
\be
\label{Vaction}
=B\abd (z-A\bd )^{-1}B\bad \psi\ad
=V\ad  (z) \psi\ad  .
\ee
It follows from~(\ref{Vaction})  that $H\ad  $  satisfies the  relation
$
H\ad  \psi\ad
=(A\ad  +V\ad  (H\ad  )) \psi\ad
$
and we can  spread  this  relation
over all the linear combinations of $H\ad  $ eigenfunctions.
Supposing that the  eigenfunctions  system  of $H\ad  $  is
dense in $\cH\ad$ we spread this
equation over ${\cal D}(A\ad  )$. As a  result
we come to the desired {\it basic equation}
(\ref{basic})
for $H\ad  $ (see also Refs.~\ct{Braun},
\ctt{MotJMPh91}{SPbWorkshop}).
Eq.~(\ref{basic}) means that the construction of the operator $H\ad  $
comes to the searching for the operator
\be
\label{Qbasic}
Q\bad
=\Int_{\sigma\bd} E\bd (d\mu )
B\bad (H\ad  -\mu )^{-1}.
\ee
Since
$
H\ad  =A\ad  +B\abd Q\bad
$,
we have
\be
\label{Qbasic1}
Q\bad =
\Int_{\sigma \bd }E\bd (d\mu )
B\bad (A\ad  +
B\abd Q\bad -\mu )^{-1},
\quad \bn \neq \an.
\ee
In this paper we restrict ourselves to the  study
of Eq.~(\ref{Qbasic1})
solvability  only  in  the  case when
spectra $\sigma_{1}$ and $\sigma_{2}$
are  separated,
\be
\label{dist}
         d_{0}={\rm dist}(\sigma _{1},\sigma _{2})>0.
\ee
Using the Lemma~\ref{LQdefinition} and the contracting  mapping  Theorem,
we prove the following:
%%%%%%%%%%%%%%%%%%%%%%%%%%%%%%%%%%%%%%%%%%%%%%%%%%%%%%%%%%%%%%%%%%%%%%%%%
\begin{theorem}\label{ThSolvability}\hspace*{-0.5em}{\sc :}
%%%%%%%%%%%%%%%%%%%%%%%%%%%%%%%%%%%%%%%%%%%%%%%%%%%%%%%%%%%%%%%%%%%%%%%%%
 Let $M\bad (\dn  )$
 be  a  set  of   bounded operators
$X,$ $X:$
            ${\cH} \ad  \rightarrow {\cH} \bd $,
 satisfying the inequality
$ \|{X}\| $ $ \le\dn     $ with $\dn  > 0$.
 If this $\dn $ and the norm
$\|{B\abd }\|_{2}$
 satisfy the condition
$
\| B\abd\|_{2}< d_{0}
\Min{} \{  \Frac{1}{1+\dn },\Frac{\dn }{1+\dn ^2}  \}$,
 then Eq.~(\ref{Qbasic1})
 is  uniquely
 solvable in
$M\abd (\dn  )$.
%%%%%%%%%%%%%%%%%%%%%%%%%%%%%%%%%%%%%%%%%%%%%%%%%%%%%%%%%%%
\end{theorem}
%%%%%%%%%%%%%%%%%%%%%%%%%%%%%%%%%%%%%%%%%%%%%%%%%%%%%%%%%%%%

\noindent{\sc Proof.} Let
\be
\label{funF}
F(X)=\Int_{\sigma\bd}E\bd(d\mu)B_{\bn\an}(A\ad+B\abd X-\mu)^{-1}
\ee
with $X$, the operator from ${\bB} (\cH\ad,\cH\bd)$.

Firstly, consider conditions when the function $F$ maps the set
$M\ad(\dn )$ into itself. We suppose here that $B\abd$ and $X$ are such
that
\be
\label{ne1}
     \|B\abd\|_2 \|X\| \leq \dn  \|B\abd\|_2 < d_0
\ee
and consequently, $\|B\abd X\|\leq d_0 . $
This means that spectrum of the operator $A\ad +B\abd X$
does not intersect with the set $\sigma_{\bn}$. Hence, the resolvent
$(A\ad +B\abd X-\mu)^{-1}$
exists and is bounded for any $\mu\in\sigma\bd$.
Thus, by Lemma~\ref{LQdefinition} we have
$$
  \|F(X)\| \leq \|B\abd\|_2\cdot E\bd\!-\!\Sup{\mu\in\sigma\bd}
  \|(A\ad +B\abd X-\mu)^{-1}\|.
$$
Due to identity
$$
(A\ad +B\abd X-\mu)^{-1}=\left( I+(A\ad-\mu)^{-1}B\abd X\right)^{-1}
  (A\ad-\mu)^{-1}
$$
and inequality $ \|B\abd\| \leq \|B\abd\|_2 $
we make estimation
$$
\| (A\ad +B\abd X-\mu)^{-1} \|  \leq
\Frac{1}{ 1 - \| (A\ad-\mu)^{-1}\| \|B\abd\|_2 \|X\| }
\|(A\ad-\mu)^{-1}\|  \leq
$$
\be
\label{estires}
\leq \Frac{1}{ 1-\Frac{1}{d_0} \|B\abd\|_2 \dn  }\cdot
\Frac{1}{d_0}=
\Frac{1}{d_0 - \|B\abd\|_2 \dn }.
\ee
Therefore, the set $M\ad(\dn )$ will be mapped by $F$ into itself if
 $\|B\abd\|_2$ and $\dn $ are such that
\be
\label{ne2}
 \|B\abd\|_2 \cdot \Frac{1}{d_0 - \|B\abd\|_2\dn } \leq \dn .
\ee

Secondly, study conditions for the function $F$ to be a contracting mapping.
Now, we consider the difference
$$
F(X)-F(Y)=\Int_{\sigma\bd} E\bd(d\mu) B\bad
\left[ (A\ad+B\abd X -\mu)^{-1} - (A\ad+B\abd Y -\mu)^{-1} \right]=
$$
$$
=\Int_{\sigma\bd} E\bd(d\mu) B\bad
 (A\ad+B\abd X -\mu)^{-1} B\abd (Y-X) (A\ad+B\abd Y -\mu)^{-1}.
$$
Again, by Lemma~\ref{LQdefinition}, we have
$$
       \|F(X)-F(Y)\|  \leq
$$
$$
\leq \|B\abd\|_2^2\cdot\Sup{\mu\in\sigma\bd}
        \| (A\ad+B\abd X -\mu)^{-1}\| \cdot
   \Sup{\mu\in\sigma\bd} \| (A\ad+B\abd Y -\mu)^{-1}\| \cdot \|(Y-X)\| .
$$
With~(\ref{estires}) we come to the estimate 
$$
\|F(X)-F(Y)\| \leq \|B\abd\|_2^2 \cdot
\Frac{1}{(d_0 -\|B\abd\|_2\dn )^2}\cdot \|Y-X\|.
$$
The function $F$ becomes a contracting mapping if
\be
\label{ne3}
 \Frac{\|B\abd\|_2^2}{(d_0 -\|B\abd\|_2\dn )^2} < 1.
\ee
Solving system of the inequalities~(\ref{ne1}),~(\ref{ne2}) and
(\ref{ne3}) we find
$$
\|B\abd\|_2 < d_0 \Min{} \left\{
\Frac{\dn }{1+\dn ^2}, \Frac{1}{1+\dn } \right\}
$$
and this completes the proof of Theorem~\ref{ThSolvability}.

\begin{corollary}\label{CBestimate}\hspace*{-0.5em}{\sc :}
Equation (\ref{Qbasic1})
is uniquely solvable in the unit
ball $M\ad(1)\subset$ ${\bB} (\cH\ad,\cH\bd)$
for any $B\abd$ such that
\be
\label{ineqB}
   \|B\abd\|_2 < \Frac{1}{2} d_0.
\ee
\end{corollary}

To prove the inequality~(\ref{ineqB}), note that
$
\Max{\dn \geq 0}\,\Min{} \left\{
\Frac{\dn }{1+\dn ^2}, \Frac{1}{1+\dn } \right\}
=\Frac{1}{2}
$
(at $\dn =1$). Hence, if~(\ref{ineqB}) takes place then the function
(\ref{funF}) is a contracting mapping of the unit ball $M\ad(1)$ into
itself.
\medskip

\noindent{\sc Remark.} In the  proofs of Lemma~\ref{LQdefinition}
and Theorem~\ref{ThSolvability}
we did not use the assumption
about finiteness of the numbers $n\ad$ of intervals included in continuous
spectra $\sigma\ad^c$ of the operators $A\ad$, $\an=1,2$. Really,
these assertions take place in the case of arbitrary spectrum
$\sigma\ad $.

Finiteness at least of one of the numbers $n_1$ and $n_2$ will be used
at the moment.
If $n_1$ and/or $n_2$ are finite and
\be
\label{BQd}
\|B\abd Q\bad\| < d_0={\rm dist}\{\sigma_1,\sigma_2\},
\quad \an=1,2,\, \bn\neq\an,
\ee
we can state  that
\be
\label{resolvest}
\|( A\ad + B\abd Q\bad -\mu )^{-1}\| \leq \Frac{C\abd}{1+|\mu|}, \,\,
\an=1,2, \quad \mbox{at any $\mu\in\sigma\bd$, $\bn\neq\an$,}
\ee
with some $C\abd > 0$, $C\abd\sim 1/(d_0 - \|B\abd Q\bad\| )$.
Of course this
estimate is essential only in the case when $\sigma\bd$ is unbounded.
It follows immediately from Eq.~(\ref{Qbasic1}) that
if $n_1$ and/or $n_2$ are finite then 
$Q\bad f\ad\in {\cal D}(H\bd)={\cal D}(A\bd)$
for any $f\ad\in \cH\ad$.

In this case we can rewrite the equation~(\ref{Qbasic1}) in  symmetric
form
\be
\label{QbasicSym}
Q\bad A\ad  -
A\bd Q\bad +
Q\bad B\abd Q\bad =B\bad .
\ee
To make  this,  it  is  sufficient  to  calculate  the
expression
$
Q\bad H\ad  -A\bd Q\bad
$
for both parts of eq.(\ref{Qbasic1}) having in mind that we apply it
to $f\ad\in {\cal D}(H\ad)$. Did, we have
$$
Q\bad H\ad  -A\bd Q\bad =
Q\bad(A\ad+B\abd Q\bad)-A\bd Q\bad=Q\bad A\ad -A\bd Q\bad +Q\bad B\abd Q\bad.
$$
On the other hand,
$$
Q\bad H\ad  -A\bd Q\bad =
\Int_{\sigma\bd}[E\bd(d\mu)B\bad(H\ad-\mu)^{-1}H\ad-
\mu E\bd(d\mu)B\bad(H\ad-\mu)^{-1}]=B\bad.
$$
One finds immediately from  Eqs.~(\ref{QbasicSym}),
$\an=1,2$, that
if $Q\bad$ gives solution $H\ad=A\ad+B\abd Q\bad$ of  the problem
(\ref{basic}) in the channel $\an$  then
\be
\label{Qadj}
Q\abd=-Q\bad^{*}=-\Int_{\sigma\ad }( H^{*}\ad
-\mu)^{-1}B\abd E\bd(d\mu )
\ee
gives analogous solution $H\ad=A\ad+B\abd Q\bad$ in the channel $\bn$.

%%%%%%%%%%%%%%%%%%%%%%%%%%%%%%%%%%%%%%%%%%%%%%%%%%%%%%%%%%%%%%%%%%%%%%%%%
\begin{theorem}\label{ThInvariant}\hspace*{-0.5em}{\sc :}
%%%%%%%%%%%%%%%%%%%%%%%%%%%%%%%%%%%%%%%%%%%%%%%%%%%%%%%%%%%%%%%%%%%%%%%%%
Let $Q\bad$, $Q\bad\in{\bB}(\cH\ad,\cH\bd)$,
be a solution of Eq.~(\ref{QbasicSym}) satisfying together with
$Q\abd=-Q\bad^{*}$ the conditions
(\ref{resolvest}). Then the transform
${\bH}'={\cal Q}^{-1}{\bH}{\cal Q}$ with
$
{\cal Q}=\left[\begin{array}{lr}   I_1      &   Q_{12} \\
                                   Q_{21}   &     I_2
\end{array}\right]
$
reduces the operator ${\bH}$ to the block--diagonal form,
${\bH}'=\diag\{ H_1, H_2 \}$ where $H\ad=A\ad + B\abd Q\bad,$
$\an,\bn=1,2,$ $\bn\neq\an.$ At the same time, the operators
$
{\cal O}\ad=\left[\begin{array}{lr}   I\ad      &     0  \\
                                      Q\bad     &     I\bd
\end{array}\right]
$
reduce the Hamiltonian ${\bH}$,
$
{\bH}=\left[\begin{array}{lr}   A\ad      &     B\abd  \\
                            B\bad     &     A\bd
\end{array}\right],
$
to triangular form,
$
{\bH}\au\equiv {\cal O}\ad^{-1} {\bH}{\cal O}\ad=
\left[\begin{array}{lr}     H\ad      &     B\abd  \\
                              0       &     H\bd^{*}
\end{array}\right].
$
\end{theorem}

\noindent{\sc Proofs} of both statements are done by direct substituting
 ${\cal Q}$ and ${\cal O}\ad$ into the definitions of ${\bH}'$ and ${\bH}\au$ and
using the equations~(\ref{QbasicSym}) .

We have to note only that operator $\cQ$ is reversible since, according to
~(\ref{Qadj}),
\be
\label{X}
X\ad=I\ad-Q\abd Q\bad = I\ad +Q\abd Q\abd^{*} \geq I\ad, \quad \an=1,2,
\ee
and
\be
\label{Qrevers}
{\cal Q}^{-1}=\left[  \begin{array}{lr}   X_1^{-1}   &   0 \\
                                                 0    &     X_2^{-1}
\end{array}\right]   \cdot
\left[  \begin{array}{lr}    I_1             &      -Q_{12} \\
                           -Q_{21}         &         I_2
\end{array}\right] .
\ee

\begin{corollary}\label{CReducing}\hspace*{-0.5em}{\sc :}
Subspaces
$\cH\au = {\cal O}  \ad(\cH\ad\oplus\{  {\bf 0} \} ) =$
$
     \{ f:\,\, f=\{ f\ad, f\bd \}\in\cH,\,\,
     f\ad\in\cH\ad,\,\, f\bd=Q\bad f\ad \}
$
are orthogonal, $  \cH^{(1)} \perp \cH^{(2)}, $
and reducing for ${\bH}$,
${\bH}\left( \cD({\bH})\bigcap\cH\au \right)\subseteq \cH\au. $
\end{corollary}

Really, if $f\in\cH\au$, $g\in\cH\bu$ and
$f=\{  f\ad, Q\bad f\ad \},$ $g=\{  Q\abd g\bd, g\bd \},$ then
$\langle f,g \rangle = $
$\langle f\ad, Q\abd g\bd \rangle + \langle  Q\bad f\ad, g\bd \rangle = 0  $
since $Q\bad = -Q\abd^{*}$.
The invariance of $\cH\au$, $\an=1,2,$ with respect to ${\bH}$ follows
 from the equality $ {\bH}\cQ =\cQ {\bH}'.$

Assertions quite analogous to the The\-o\-rem \ref{ThInvariant}
and Corol\-la\-ry \ref{CReducing} one
can find in Refs.~\ct{MalyshevMinlos1},   {}\ct{MalyshevMinlos2}.
Solvability (for sufficiently small $\|B\abd\|$)
of the equation~(\ref{QbasicSym}) was proved
in\ct{MalyshevMinlos1},\ct{MalyshevMinlos2} by rather different method
also in the supposition~(\ref{dist}).
\medskip

\noindent{\sc Remark}.  It follows from
Theorem~\ref{ThInvariant}  that operator
$  \tilde{\cQ}=\cQ X^{-1/2} $ with $X=\diag\{X_1,X_2\}$
is unitary. Consequently, the operator
 $ {\bH}''= \tilde{\cQ}^{*} {\bH} \tilde{\cQ}=$ $X^{1/2} {\bH}' X^{-1/2}$
becomes self-adjoint in $\cH $.
Since $ {\bH}''=\diag\{  H''_1, H''_1  \}$
with $ H''\ad=X\ad^{1/2} H\ad X\ad^{-1/2}$,
 the operators $H''\ad$, $\,\,\,\an=1,2,$
are self-adjoint on $\cD(A\ad)$
in $\cH\ad $. Moreover the operators
$
{\bH}\au=\tilde{\cQ}\cdot \diag\{ H''\ad, 0\}\cdot  \tilde{\cQ}^{*} =
$
$\cQ \cdot\diag\{ H\ad, 0 \}\cdot \cQ^{-1} $  represent
parts of the Hamiltonian ${\bH}$ in the corresponding invariant subspaces
$\cH^{(1)}$ and $\cH^{(2)}$
(see also Refs.~\ct{MalyshevMinlos1},\ct{MalyshevMinlos2}).

Unfortunately, eigenvectors ${{\psi}''\ad}$
of the operators $H''\ad$  differ from those
for the initial spectral problem~(\ref{ini}):
${{\psi}''}\ad=X\ad^{1/2}\psi\ad$.

%%%%%%%%%%%%%%%%%%%%%%%%%%%%%%%%%%%%%%%%%%%%%%%%%%%%%%%%%%%%%%%%%%%%%%%%%%%%
\begin{lemma}\label{LHolder}\hspace*{-0.5em}{\sc :}
%%%%%%%%%%%%%%%%%%%%%%%%%%%%%%%%%%%%%%%%%%%%%%%%%%%%%%%%%%%%%%%%%%%%%%%%%%%%
 Let the kernel $B\bad(\mu,\lambda)$, $\bn\neq\an$,
of the operator $B\bad$ belong to the class $\cB^{\bn\an}_{\theta\gamma}$
with $\theta > \Frac{1}{2}$
and $Q\bad$ be a solution of Eq.~(\ref{Qbasic1}) satisfying together
with $Q\abd=-Q\bad^{*}$ the conditions
(\ref{resolvest}). Then

(a) the operator $Q\bad$ is an integral operator, $Q\bad:$
$\cH\ad\rightarrow\cH\bd,$ with a kernel $Q\bad(\mu,\lambda)$
belonging to $\cB^{\bn\an}_{\theta\gamma}$;

(b) the potential $W\ad\equiv B\abd Q\bad$ is
an integral operator, $W\ad:$
$\cH\ad\rightarrow\cH\ad,$ with a kernel $W\ad(\lambda,\lambda')$
belonging to $\cB^{\an\an}_{\theta\gamma}$.
\end{lemma}

\noindent{\sc Proof.} At the beginning we prove the assertion (b).
According to~(\ref{Qadj}),
\be
\label{w}
W\ad=-B\abd\Int_{\sigma\ad}(H\bd^{*}-\lambda)^{-1}B\bad E\ad(d\lambda)
\ee
with $H\bad^{*}=A\bd+W\bd^{*}=A\bd+Q\abd^{*}B\abd$. Since the inequalities
(\ref{resolvest}) take place we write
$$
\| (H\bd^{*}-\lambda)^{-1} \| = \| (H\bd-\lambda)^{-1} \|
\leq C\bad
$$
for any $\lambda\in\sigma\ad$. In the diagonal representation
(\ref{Neumann}),(\ref{multi}), the equation~(\ref{w}) turns in
$$
W\ad(\lambda,\lambda')=-B\abd(\lambda,\,\cdot\,)(H\bd^{*}-\lambda)^{-1}
B\bad(\,\cdot\, ,\lambda').
$$
It means that
$$
|W\ad(\lambda,\lambda')| \leq
\|B\abd(\lambda,\,\cdot\, )\|_{\cH\bd}\cdot
\|(H\bd^{*}-\lambda)^{-1}\|\cdot
\| B\bad(\,\cdot\, ,\lambda')\|_{\cH\bd}\leq
$$
\be
\label{West}
\leq C\bad \|B\abd(\lambda,\,\cdot\, )\|_{\cH\bd}\cdot
\| B\bad(\,\cdot\, ,\lambda')\|_{\cH\bd}.
\ee
Here,
$
\|B\abd(\lambda,\,\cdot\, )\|_{\cH\bd}=
\left[  \intsum{\sigma\bd} |B\abd(\lambda,\mu)|^{2}d\mu \right]^{1/2}.
$
Since $\theta >\Frac{1}{2},$ we have
$
\|B\abd(\lambda,\,\cdot\, )\|_{\cH\bd}
$ $\leq$ $ \Frac{c(\theta)}{(1+|\lambda|)^\theta}\cdot \|B\|_{\cB}
$
with some $c(\theta)$, $\, c(\theta) > 0$, depending only on $\theta$.
Analogously,
$$
\|B\bad(\,\cdot\, ,\lambda')\|_{\cH\bd}=
\|\overline{B\abd}(\lambda',\,\cdot\,)\|_{\cH\bd}
\leq \Frac{c(\theta)}{(1+|\lambda|)^\theta}\cdot \|B\|_{\cB},
$$
where the operator
$\overline{B\abd}(\lambda,\mu),$ $\overline{B\abd}(\lambda,\mu):$
$\cG\bd(\mu)\rightarrow\cG\ad(\lambda)$,
is adjoint to $B\bad(\mu,\lambda)$.

Estimations similar to~(\ref{West}) may be done also for
$|W\ad(\lambda'',\lambda')-W\ad(\lambda,\lambda')|,$
$\, \lambda,\lambda''\in\sigma\ad^c,$
$\, \lambda\in\sigma\ad$,
$|W\ad(\lambda,\lambda''')-W\ad(\lambda,\lambda')|,$
$\, \lambda\in\sigma\ad$,
$\, \lambda''',\lambda'\in\sigma\ad^c,$
and
$ |W\ad(\lambda,\lambda')-W\ad(\lambda'',\lambda') -$
$W\ad(\lambda,\lambda''')+W\ad(\lambda'',\lambda''')|,\,$
$\, \lambda,\lambda',\lambda'',\lambda'''\in\sigma\ad^c$,
in terms of the norms
$
\|B\abd(\lambda,\,\cdot\,) - B\abd(\lambda'',\,\cdot\,)\|_{\cH\bd}
$
                and
$
\|B\bad(\,\cdot\, ,\lambda''') - B\bad(\,\cdot\, ,\lambda')\|_{\cH\bd}.
$
Estimating the latter through $ \| B\abd\|_\cB $
we come to the inequality
$$
  \| W\ad\|_{\cB^{\an\an}_{\theta\gamma}}
  \leq c(\theta)C\bad\cdot
       \| B\bad\|^2_{\cB^{\bn\an}_{\theta\gamma}}
$$
with  $0< c(\theta) < \infty. $  Therefore, we have proved the assertion (b).

To prove the statement (a) we note that according to~(\ref{Qbasic1}),
$$
Q\bad=\Int_{\sigma\bd} E\bd(d\mu) B\bad \left[
(A\ad -\mu)^{-1}-(H\ad-\mu)^{-1}W\ad(A\ad-\mu)^{-1}  \right]
$$
or, in the diagonal representation~(\ref{Neumann}),(\ref{multi}),
$$
Q\bad(\mu,\lambda)=\Frac{B\bad(\mu,\lambda)}{\lambda-\mu}
- \Frac{B\bad(\mu,\,\cdot\,)(H\ad-\mu)^{-1}W\ad(\,\cdot\, ,\lambda)}
{\lambda-\mu}.
$$
Repeating literally the last part of the proof of the assertion (b)
we come to the inequality
$$
\| Q\bad\|_{\cB^{\bn\an}_{\theta\gamma}} \leq
\Sup{
                           \mbox{
                                  \scriptsize
                                           $\begin{array}{c}
                                                 \mu\in\sigma\bd\\
                                             \lambda\in\sigma\ad
                                             \end{array}$
                                 }
    }
     \Frac{1}{   |\lambda-\mu| } \cdot
                                                 \left\{
     \| B\bad \|_{  \cB^{\bn\an}_{\theta\gamma}   } +
                    \mbox{\phantom{ $\Sup{\mu\in\sigma\bd}$ }}  \right.
$$
$$
                                          \left. \mbox{\phantom{{\Large I}}}
+ c(\theta)\cdot\| B\bad\|_{\cB^{\bn\an}_{\theta\gamma}} \cdot
\Sup{\mu\in\sigma\bd}  \| (H\ad-\mu)^{-1}  \|\cdot
\| W\ad\|_{\cB^{\an\an}_{\theta\gamma}}
                                                  \right\},
               \quad 0 < c(\theta) < +\infty.
$$
Consequently $Q\bad\in\cB^{\bn\an}_{\theta\gamma} $
and
$$
\|Q\bad\|_{\cB} \leq \Frac{1}{d_0}\cdot
                                                     \left\{
\|B\bad\|_{\cB}  +
c(\theta)C\abd C\bad
\cdot \|B\bad\|_{\cB}^3                               \right\},
\quad 0 <c(\theta) < +\infty.
$$
This completes the proof of Lemma~\ref{LHolder}.

\noindent\begin{corollary}\label{CHolderSmoothness}\hspace*{-0.5em}{\sc :}
{\it If $B\bad\in\cB^{\bn\an}_{\theta\gamma}$,
$\,\theta > \Frac{1}{2},$
then the solution of Eq.~(\ref{Qbasic1}) described by
Theorem~\ref{ThSolvability}
belongs to the class $\cB^{\bn\an}_{\theta\gamma}$, too.   }
\end{corollary}

This statement is based on the fact that the mentioned solution satisfies
automatically the conditions~(\ref{BQd}) and, hence, the conditions
(\ref{resolvest}).
%%%%%%%%%%%%%%%%%%%%%%%%%%%%%%%%%%%%%%%%%%%%%%%%%%%%%%%%%%%%%%%%%%%%%%%%%%%%
\section{\hspace*{-1em}. EIGENFUNCTIONS AND THE EXPANSION THEOREM}
\label{Eigenfunctions}
%%%%%%%%%%%%%%%%%%%%%%%%%%%%%%%%%%%%%%%%%%%%%%%%%%%%%%%%%%%%%%%%%%%%%%%%%%%%
In the preceding section, we have proved the existence
(in the unit ball $M\ad(1)\subset\cH\ad$)
of a solution
$Q\bad$ of the basic equation~(\ref{Qbasic1}) only in the case when spectra
$\sigma_1,$ $\sigma_2$ of the operators $A_1$, $A_2$ are separated,
${\rm dist}\{  \sigma_1,\sigma_2\}$ = $d_0$ $> 0$, and
$\|B_{12}\|_2 = \| B_{21} \|_2 < \Frac{d_0}{2}$. May be, however,
Eqs.~(\ref{basic}) and~(\ref{Qbasic1}) have solutions also in other
cases. That is why we study the spectral properties of the operator
$H\ad=A\ad+B\abd Q\bad$ not supposing that
$\|B\abd\|_2 < \Frac{d_0}{2}$
and using more general requirements~(\ref{resolvest}) only, with
$C\abd$,  some positive numbers, $\an,\bn=1,2$, $\bn\neq\an$. Of course, we
assume again that the condition~(\ref{dist}) takes place.
Remember that the requirements~(\ref{resolvest}) are sufficient for
existence of the operators $V\ad(H\ad)$. As well, the equations
(\ref{QbasicSym}) and~(\ref{Qadj}) take place and the assertions of
Theorem~\ref{ThInvariant} and Lemma~\ref{LHolder} are valid.

So, let us suppose that $Q\bad$ and $Q\abd=-Q\bad^{*}$ are solutions
of Eqs.~(\ref{Qbasic1}) and~(\ref{QbasicSym}) satisfying the
conditions~(\ref{resolvest}). It follows from Lemma~\ref{LQdefinition} that
$Q\bad\in\bB\bad(\cH\ad,\cH\bd)$ as well as $Q\abd\in\bB\abd(\cH\bd,\cH\ad)$.
If $B\bad\in\cB_{\theta\gamma}^{\bn\an}$, $\theta>\Frac{1}{2}$, then,
according to Lemma~\ref{LHolder}, $Q\bad\in\cB_{\theta\gamma}^{\bn\an}$
and $Q\abd\in\cB_{\theta\gamma}^{\an\bn}$ .

By Theorem~\ref{ThInvariant}, the operator
${\bH}'=\diag\{ H_1, H_2  \}$ is connected with
the (self-adjoint) operator  ${\bH}$
by a similarity transform. Thus, the spectra
$\sigma(H_1)$ and $\sigma(H_2)$ of the operators $H\ad$, $\an=1,2$,
are real and $\sigma(H_1)\bigcup\sigma(H_2)=\sigma({\bH})$. Continuous spectrum
$\sigma_c (H\ad)$ of the each operator $H\ad$ coincides with that of the
operator $A\ad$, $\sigma_c (H\ad)=\sigma\ad^c$,
since due to $\|B\abd\|_2 < +\infty$, the potential $W\ad=B\abd Q\bad $ is
a compact operator. Since $\sigma_1^c \bigcap\sigma_2^c =\emptyset $
we have $\sigma_c (H_1)\bigcap\sigma_c (H_2) =\emptyset$. We show now that the
discrete spectra $\sigma_d (H\ad),$ $\an=1,2$, satisfy a similar condition.

Let us suppose that $\sigma _{d}(H\ad  )\neq\emptyset $,
$z\in \sigma _{d}(H\ad  )$ and $\psi \ad  $  is
the  corresponding  eigenfunction  of $H\ad  ,$
$ H\ad  \psi \ad  =z\psi \ad  ,$
$\psi \ad  \in {\cal D}(H\ad  )={\cal D}(A\ad  )$.
Then, according to construction  of $H\ad$,
we   have
$H\ad  \psi \ad  =$
$(A\ad  +V\ad  (H\ad  ))\psi \ad  =$
$(A\ad  +V\ad  (z))\psi \ad  =z\psi \ad  $.
Thus if $z\in \sigma _{d}(H\ad  )$  then $z$  becomes
automatically a point  of the discrete
spectrum of the  initial spectral problem ~(\ref{ini}).
At the same time $\psi\ad $ becomes it's eigenfunction.

Let us further denote the eigenfunctions  of  the
operator $H\ad  $  discrete  spectrum  by $\psi \ju \ad  , $
$\psi \ju \ad  =u\ju \ad  $,
keeping for them the  same  numeration  as  for
eigenvectors of $U_j$, $U_j=\{ u\ad\ju, u\bd\ju \}, $
of the Hamiltonian ${\bH}$,
${\bH}U_j =z_j U_j$, $z_j\in\sigma_d ({\bH}).$ We assume that in the case
of multiple discrete eigenvalues, certain $z_j$ may be repeated
in this numeration.
 By $\cU ^{d}$
we denote the set $\cU ^{d}=\{U_{j}, j=1,2,\ldots\}$
of  all the eigenvectors $U_j$.

Let $\cU ^{d}\ad  $
be such a subset of $\cU ^{d}$
that it's elements have
the operator $H\ad  $ eigenvectors $\psi \ju \ad  $
in the  capacity  of the  channel $\an $
components:
$\cU \ad^d =$ $\{U_{j}:$
$ U_{j}=\{u\ju _{1},u\ju _{2}\},$
$ u\ju \ad  =\psi \ju \ad  \}$.
By Theorem~\ref{ThInvariant}, we have
$\cU ^{d}_1  \bigcup \cU ^{d}_2 = \cU ^{d}.$

%%%%%%%%%%%%%%%%%%%%%%%%%%%%%%%%%%%%%%%%%%%%%%%%%%%%%%%%%%%%%%%%%%%%%
\begin{theorem}\label{ThDiscSpectrum}\hspace*{-0.5em}{\sc :}
%%%%%%%%%%%%%%%%%%%%%%%%%%%%%%%%%%%%%%%%%%%%%%%%%%%%%%%%%%%%%%%%%%%%%
Let $H\bd  =A\bd  + B\bad Q\abd ,$
correspond  (for
$\| B\bad \|_{2} < +\infty $)
to the same  solution $Q\abd =- Q\bad^{*} $
of  Eqs.~(\ref{Qbasic1}) and~(\ref{QbasicSym})  as
$H\ad =A\ad +B\abd Q\bad $,
and the conditions~(\ref{resolvest}) are valid.
Let $z_{j}\in \sigma _{d}(H\ad )$
 and $H\ad u\ju \ad =z_{j}u\ju \ad $ with $u\ad\ju $,
the channel $\an$ component of the eigenvector
$U_j =\{ u\ad\ju,u\bd\ju \}$ of the operator ${\bH}$, ${\bH}U_j =z_j U_j$.
Then either
 $z_{j}\not\in \sigma _{d}(H\bd)$, $\bn\neq\an$,
 or  (if $z_{j}\in\sigma_{d}(H\bd  ))$ the vector $u\bd\ju $ is not
an eigenvector of $H\bd$.
\end{theorem}

\begin{corollary}\label{CDiscSpectrum}\hspace*{-0.5em}{\sc :}
$
\cU ^{d}_1
\bigcap\cU ^{d}_2 =\emptyset
$.
\end{corollary}

Statement of Theorem~\ref{ThDiscSpectrum} means that discrete
spectrum $\sigma_d ({\bH}) $ is distributed between
discrete spectra $\sigma_d (H_1)$ and $\sigma_d (H_2)$ in such a way that
operators $H_1$ and $H_2$ have not ``common'' eigenvectors
$U_j =\{ u_1\ju,u_2\ju \}$: simultaneously, component $u_1\ju$
can not be eigenvector
for $H_1$, and  $u_2\ju$ with the same $j$,  for $H_2$.
\medskip

\noindent{\sc Proof} of the Theorem will be given by contradiction.

Let us suppose that $\psi\ad\ju = u\ad\ju $ is an eigenvector of $H\ad$
corresponding to $z_j$ i.e.
\be
\label{Aeigen}
(A\ad+B\abd Q\bad -z_j)\psi\ad\ju=0.
\ee

If $z_j\in\sigma\ad=\sigma(A\ad)$ then automatically
$z_j\not\in\sigma_d(H\bd)$ since due to conditions~(\ref{resolvest})
we have $\sigma(H\bd)\bigcap\sigma(A\ad)=\emptyset$. Thus in the case
when $z_j\in\sigma\ad$ the assertion of Theorem is valid.

Let $z_j\not\in\sigma(A\ad)$. In this case we can rewrite
Eq.~(\ref{Aeigen}) in the form
\be
\label{LSch}
\psi\ad\ju = -(A\ad -z_j)^{-1}B\abd Q\bad\psi\ad\ju.
\ee
Let $y\bd\ju=Q\bad\psi\ad\ju$. It follows from~(\ref{LSch}) that
\be
\label{y}
y\bd\ju + Q\bad(A\ad - z_j)^{-1} y\bd\ju =0.
\ee
We will show that the vector $ y\bd\ju $ is a solution of the initial
spectral problem~(\ref{ini}) in the channel $\bn$ at $z=z_j$ and
$\tilde{U}_j=\{ \psi\ad\ju, y\bd\ju \}$ is an eigenvector of ${\bH}$,
$\bH\tilde{U}_j=z_j \tilde{U}_j$. To do this, we act on both parts of
Eq.~(\ref{y}) by    $   H^{*}\bd-z_j  $  remembering that, according to
(\ref{Qadj}),
$Q\bad=-Q\abd^{*}$
$   =-\Int_{\sigma\ad}(H\bd^{*}-\lambda)^{-1}B\bad E\ad(d\lambda)    $.
We obtain
$$
          (H\bd^{*}-z_j)y\bd\ju+\Int_{\sigma\ad} (H\bd^{*}-z_j)
      (H\bd^{*}-\lambda)^{-1}(z_j -\lambda)^{-1} B\bad E\ad(d\lambda)
                        B\abd y\bd\ju =0.
$$
Using the identity
   $       (H-z)(H-\lambda)^{-1}(z-\lambda)^{-1}=
            (z-\lambda)^{-1}-(H-\lambda)^{-1} $
we find
$$
          (H\bd^{*}-z_j)y\bd\ju+\Int_{\sigma\ad}
      [(z_j -\lambda)^{-1} - (H\bd^{*}-z_j) ]
       B\bad E\ad(d\lambda) B\abd y\bd\ju =0
$$
or, and it is the same,
\be
\label{yy}
  (H\bd^{*}-z_j)y\bd\ju - B\abd(A\ad -z_j)^{-1} B\abd y\bd\ju
   + Q\bad B\abd y\bd\ju =0.
\ee
However $ H\bd^{*}=A\bd -Q\bad B\abd $. Hence the relation~(\ref{yy})
turns in equation~(\ref{ini}) for the channel $\bn$,
$$
  [A\bd - B\bad (A\ad -z_j)^{-1} B\abd - z_j ] y\bd\ju = 0.
$$
So, we have proved that $y\bd\ju$ is a solution of the initial problem
in the channel $\bn$ and we did deal with an eigenvector
$ U_j=\{ u\ad\ju, u\bd\ju \} $ of the operator ${\bH}$ having the components
$u\ad\ju=\psi\ad\ju$ and $u\bd\ju =y\ad\ju $.

Let us show that $y\bd\ju$ can not be an eigenvector of $H\bd$
corresponding to the eigenvalue $z_j$. Actually, due to~(\ref{y}) we have
$$
{\rm a}\equiv \langle y\bd\ju +Q\bad (A\ad -z_j)^{-1}B\abd y\bd\ju,
 y\bd\ju  \rangle  =0.
$$
On the other hand
$$
{\rm a}= \| y\bd\ju\|^2 + \langle (A\ad-z_j)^{-1} B\abd y\bd\ju,
  Q\bad^{*}y\bd\ju   \rangle.
$$
If $y\bd\ju$ is an eigenvector of $H\bd$, $H\bd y\bd\ju =z_j y\bd\ju$,
then
$$
Q\bad^{*} y\bd\ju =-Q\abd y\bd\ju = -\Int_{\sigma\ad}
E\ad(d\lambda) B\abd (H\bd -\lambda)^{-1} y\bd\ju =
  (A\ad -z_j)^{-1} B\abd y\bd\ju.
$$
It means that
$$
{\rm a} = \|y\bd\ju\|^2 +\| (A\ad-z_j)^{-1} B\abd y\bd\ju\|^2
\geq  \|y\bd\ju\|^2 .
$$
Since ${\rm a} =0 $ we get  $y\bd\ju =0 $
and, due to~(\ref{LSch}), $\psi\ad\ju =0.$ However, by supposition,
$ \psi\ad\ju \neq 0.$
 Thus, we come to a contradiction and $y\bd\ju$ can not be
an eigenvector of $H\bd$. And so, if $z_j \in \sigma_d (H\ad)$ and
$H\ad u\ad\ju =z_j u\ad\ju $ then $u\bd\ju $ is not an eigenvector of $H\bd$.
The proof of Theorem~\ref{ThDiscSpectrum} is completed.
\medskip

Let us pay attention to the continuous spectrum of $H\ad$
assuming here that $B\abd\in\cB^{\bn\an}_{\theta\gamma},$
$\theta > \frac{1}{2},$ $\gamma > \frac{1}{2}$, and consequently,
$Q\abd\in\cB^{\bn\an}_{\theta\gamma},$ $\an,\bn=1,2$, $\bn\neq\an.$

Consider at $\lambda'\in\sigma\ad^c$ the integral equations
\be
\label{LSchpsi}
\psi ^{(\pm )}\ad  (\lambda ,\lambda ')=
I\ad^c  \dn  (\lambda -\lambda ')-
[(A\ad  -\lambda '\mp i0)^{-1}
W\ad
\psi ^{(\pm )}\ad  ](\lambda ,\lambda '), \,\, \an=1,2,
\ee
where as usually $W\ad=B\abd Q\bad. $ Since
$W\ad\in\cB^{\an\an}_{\theta\gamma},$
the integral operator with the kernel
$
\Frac{W\ad(\lambda,\lambda')}{\lambda -\lambda '\mp i0}
$
is compact  in
$
{\cal M}_{\theta '\gamma '},
$
$
\Frac{1}{2}<\theta '<\theta ,
$
$
0<\gamma '<\gamma
$
(cf. Refs.~\ct{LadyzhFaddeev},\ct{Faddeev64}).
If $\lambda '\not\in \sigma _{d}(H\ad  )$
then Eq.~(\ref{LSchpsi}) for
$\psi^{(+)}_{\an}$
as well as for $\psi^{(-)}_{\an}$ is
uniquely  solvable (see Ref.~\ct{Faddeev64}) in
the  class  of  the  form ~(\ref{amplitude})
distributions.

Denote by
$
\Psi ^{(\pm )}\ad  ,
$
$\Psi ^{(\pm )}\ad  :$
$
{\cH} ^{c}\ad  \rightarrow {\cH} \ad
$,
the in\-te\-gral ope\-ra\-tor
with the  ker\-nel
$
\psi ^{(\pm )}\ad  (\lambda ,\lambda ')
$.
The  operator $\Psi ^{(\pm )}\ad  $  is
bounded and
$
\Psi ^{(\pm )}\ad  {\cal D}(A^{(0)}\ad  )
\subseteq {\cal D}(H\ad  )
$~\ct{LadyzhFaddeev},\ct{Faddeev64}.
It  follows  from ~(\ref{LSchpsi})
that $\Psi ^{(\pm )}\ad  $
has   the   property
$
H\ad  \Psi ^{(\pm )}\ad  =
\Psi ^{(\pm )}\ad  A^{(0)}\ad
$.
Thus,
$
Q\bad \Psi ^{(\pm )}\ad  (\cdot ,\lambda ')=
$
$
(\lambda '- A\bd )^{-1}
$
$
B\bad \Psi ^{(\pm )}\ad  (\cdot ,\lambda ')
$.
Substitution    of    this
expression in~(\ref{LSchpsi}) shows that
$
\psi ^{(\pm )}\ad
$
satisfies~(\ref{wavefunction}).
Due to  the  uniqueness
of  Eq.~(\ref{wavefunction}) solution at $\lambda'\not\in\sigma_d({\bH})$
  we   have
$
\psi ^{(\pm )}\ad  (\lambda ,\lambda ')=
u^{(\pm )}_{\an\an}(\lambda,\lambda')
$.
This    means    that    each
eigenfunction
$
u^{(\pm )}_{\an \an }(\lambda ,\lambda '),
$
$
\lambda ^\prime \in \sigma ^{c}\ad  ,
$
$
\lambda '\not\in \sigma _{d}({\bH})
$
of the initial spectral problem
problem~(\ref{ini}) is also an eigenfunction  of $H\ad  $.

Consider  the  functions
$
\tilde{\psi}\ju \ad  =
\psi \ju \ad  -Q\abd u\ju \bd
$
and \newline
$
\tilde{\psi}^{(\pm )}\ad  (\cdot ,\lambda ')=
$
$
\psi ^{(\pm )}\ad  (\,\cdot\, ,\lambda ')-
$
$
Q\abd u^{(\pm )}\bad
(\,\cdot\, ,\lambda '),
$
$
\lambda '\in \sigma ^{c}\ad
$.
Let
$
\tilde{\Psi }^{(\pm )}\ad  ,
$
$
\tilde{\Psi }^{(\pm )}\ad  :
$
$
{\cH} ^{c}\ad  \rightarrow {\cH} \ad
$,
be the integral operator  with  the
kernel
$
\tilde{\psi}^{(\pm )}\ad  (\lambda ,\lambda ')
$.

%%%%%%%%%%%%%%%%%%%%%%%%%%%%%%%%%%%%%%%%%%%%%%%%%%%%%%%%%%%%%%%%%%%%%%%%%
\begin{theorem}\label{ThExpansion}\hspace*{-0.5em}{\sc :}
%%%%%%%%%%%%%%%%%%%%%%%%%%%%%%%%%%%%%%%%%%%%%%%%%%%%%%%%%%%%%%%%%%%%%%%%%
 The functions
$
\tilde{\psi}\ju \ad
$
(with $j$  such  that
$
U_{j}\in \cU ^{d}\ad  )
$
 are eigenfunctions of adjoint  operator
$
H^{*}\ad  , H^{*}\ad  =
A\ad  +Q^{*}\bad B\bad ,
$
  discrete  spectrum,
$
H^{*}\ad  \tilde{\psi}\ju \ad  =
$
$
z_{j}\tilde{\psi}\ju \ad
$.
 Operators
$
\tilde{\Psi }^{(\pm )}\ad
$
 have the property
$
H^{*}\ad  \tilde{\Psi }^{(\pm )}\ad=
$
$
\tilde{\Psi }^{(\pm )}\ad  A^{(0)}\ad
$.
 At the same  time  the  orthogonality  relations  take
place:
$
\langle\psi\ju \ad  ,\tilde{\psi}^{(k)}\ad  \rangle=
\dn  _{jk},
$
$
\Psi ^{(\pm )*}\ad  \tilde{\Psi }^{(\pm )}\ad  =\reduction{I\ad}{\cH\ad^c},
$
$
\tilde{\Psi }^{(\pm )*}\ad  \psi \ju \ad  =0
$
                       and
$
\Psi\ad^{(\pm)*}\tilde{\psi}\ju \ad  =0.
$
  Also,  the following  completeness  relations are valid,
\be
\label{completeness}
\Sum_{j:\, U_{j}\in\cU \ad^d}
{\psi}\ju \ad \langle\cdot,{\psi}\ju \ad  \rangle+
\Psi^{(\pm )}\ad  \tilde{\Psi }^{(\pm )*}\ad  =I\ad  , \quad \an =1,2,
\ee
\end{theorem}

\noindent{\sc Proof.} Show for example that
\be
\label{Hadjeigen}
  H\ad^{*}\tpsi\ad\ju=z_j\tpsi\ad\ju
\ee
(remember that $z_j\in\bR$). We have
$$
H\ad^{*}\tpsi\ad\ju=(A\ad-Q\abd B\bad)(\tpsi\ad\ju -Q\abd u\bd\ju)=
$$
\be
\label{Heq1}
=(A\ad-Q\abd B\bad)\psi\ad\ju -(A\ad Q\abd-Q\abd B\bad Q\abd)u\bd\ju.
\ee
Note that $A\ad=H\ad-B\abd Q\bad$ and, hence,
\be
\label{Heq2}
(A\ad-Q\abd B\bad)\psi\ad\ju =z_j\psi\ad\ju -
(B\abd Q\bad+Q\abd B\bad)\psi\ad\ju.
\ee
Second term in the right part of~(\ref{Heq2}) may be easily expressed through
$u\bd\ju$. Actually,
$ u\bd\ju=-(A\bd-z_j)^{-1}B\bad \psi\ad\ju $
(we use again the property $\sigma(H\ad)\bigcap\sigma\bd=\emptyset$
following from~(\ref{resolvest}) ). Since Eqs.~(\ref{Qbasic}) and
$H\ad\psi\ad\ju=z_j\psi\ad\ju$
take place, we find $Q\bad\psi\ad\ju=B\abd u\bd\ju.$
Consequently,
$$
           (B\abd Q\bad +Q\abd B\bad)\psi\ad\ju=
    B\bad Q\bad\psi\ad\ju+Q\abd(A\bd-z_j)(A\bd-z_j)^{-1}B\bad\psi\ad\ju=
$$
$$
         =B\bad u\bd\ju -Q\abd (A\bd - z_j)u\bd\ju.
$$
Substituting the expressions obtained into~(\ref{Heq2}) and then
into~(\ref{Heq1}), we get
$$
            H\ad^{*}\tpsi\ad\ju=z_j(\psi\ad\ju-Q\abd u\bd\ju)
           + [-B\abd+Q\abd A\bd- A\ad Q\abd +Q\abd B\bad Q\abd]u\bd\ju.
$$
According to the equations~(\ref{QbasicSym}), the expression in the square
brackets is equal to zero and we come to~(\ref{Hadjeigen}).

The equalities
$
     H\ad^{*}\tpsipm\ad(\,\cdot\, ,\lambda')=
$
$
     \lambda'\tpsipm\ad(\,\cdot\, ,\lambda'),
$
$\lambda'\in\sigma\ad^c ,$ are proved quite analogously.

The orthogonality relations
$
\langle\psi\ju \ad  ,\tilde{\psi}^{(k)}\ad  \rangle=
\dn  _{jk},
$
$
\tilde{\Psi }^{(\pm )*}\ad  \psi \ju \ad  =0
$
                       and
$
\Psi\ad^{(\pm)*}\tilde{\psi}\ju \ad  =0
$
are trivial. Proofs of the relation
$
\Psi ^{(\pm )*}\ad  \tilde{\Psi }^{(\pm )}\ad  =\reduction{I\ad}{\cH\ad^c},
$
and the equality~(\ref{completeness}) are very similar. Both these proofs are
based on use of properties of the wave operators $U^{(\pm)}$. As a sample,
we give a proof of the completeness relation~(\ref{completeness}).

Consider the operator
$$
{\cal A}=\Sum_{j:\, U_j\in\cU\ad^d}
\psi\ad\ju \langle\,\cdot\, ,\tpsi\ad\ju\rangle +
\Psi\ad\tPsi\ad^{*} =
$$
\be
\label{Aeq0}
=\Sum_{j:\, U_j\in\cU\ad^d}
\psi\ad\ju \langle\,\cdot\, ,\psi\ad\ju -Q\abd u\bd\ju\rangle +
\Psi\ad[\Psi\ad^{*} -(Q\abd u\bad)^{*}].
\ee
For convenience, we omit signs ``$\pm$'' in notations of
$\Psipm\ad \equiv  u^{(\pm)}\aad$,
$ u^{(\pm)}\bad$
and
$\tPsipm\ad$
taking in mind for example the case of sign ``$+$''. We have from
(\ref{Aeq0}):
$$
{\cal A}=\Sum_{j:\, U_j\in\cU\ad^d}
\psi\ad\ju \langle\,\cdot\, ,\psi\ad\ju   \rangle +
\Psi\ad \Psi\ad^{*} -
\Sum_{j:\, U_j\in\cU\ad^d}
\psi\ad\ju \langle\,\cdot\, , Q\abd u\bd\ju  \rangle
 -\Psi\ad  u\bad^{*} Q\abd^{*}.
$$
It follows from the completeness relations $U^{(\pm)*}U^{(\pm)}=I-P$
for wave operators $U^{(\pm)}$ that
\be
\label{Aeq1}
            \Psi\ad u\bad^{*}\equiv  u\aad u\bad^{*}=
         -u\abd u\bbd^{*} -\Sum_{z_j\in\sigma({\bH})}
      u\ad\ju     \langle    \,\cdot\, ,u\bd\ju     \rangle.
\ee
Since $ u\bbd^{*} Q\abd^{*} =(Q\abd u\bbd)^{*} = u\abd^{*} ,$ we can write
with a help of~(\ref{Aeq1}) that
$$
    {\cal A}=u\aad u\aad^{*} + u\abd u\abd^{*} +
    \Sum_{j:\, U_j\in\cU\ad^d}
    \psi\ad\ju \langle \,\cdot\, ,\psi\ad\ju   \rangle +
\Sum_{                             \mbox{
\scriptsize
$\begin{array}{c}
          z_j\in\sigma_d (\mbox{{\footnotesize$\bH$}})\\
           U_j \not\in \cU\ad^d
\end{array}$
                                         }
    }
     u\ad\ju  \langle \,\cdot\, , Q\abd u\bd\ju   \rangle .
$$
In the last sum, the conditions
$   z_j\in\sigma_d ({\bH})  $  and   $  U_j \not\in \cU\ad^d  $
mean really that we deal with any $j$ such that
$U_j\in\cU\bd^d$.   This follows from the equalities
$ \cU_1^d \bigcup \cU_2^d = \cU^d $  and
$ \cU_1^d \bigcap \cU_2^d =  \emptyset $  (see
Theorem~\ref{ThDiscSpectrum} and Corollary~\ref{CDiscSpectrum}).
For $ U_j\in\cU\bd^d$, the vector $u\bd\ju$ is eigenfunction of $H\bd$,
$u\bd\ju=\psi\bd\ju$, and $Q\abd u\bd\ju =$ $Q\abd \psi\bd\ju = u\ad\ju .$
Thus, ${\cal A}$ turns in
$$
{\cal A}=u\aad u\aad^{*} + u\abd u\bad^{*}
+ \Sum_{z_j \in \sigma_d ({\bH})}
     u\ad\ju  \langle \,\cdot\, ,  u\ad\ju   \rangle
     =(U^{(\pm)}U^{(\pm)*} + P)\aad.
$$
Since $U^{(\pm)}U^{(\pm)*}+P=I$ we find
 ${\cal A}=I\ad$ and this completes the proof of Theorem~\ref{ThExpansion}.
\medskip

Theorem~\ref{ThExpansion} means in particular that  {\it
 part $H^{c}\ad  $ of
operator $H\ad  $   acting   in   the   invariant   subspace
corresponding to it's continuous spectrum $\sigma\ad^c$,  is  similar
to the operator
$
A^{(0)}\ad,
$
$
H^{c}\ad  =
\Psi ^{(\pm )}\ad  A^{(0)}\ad
\tilde{\Psi }^{(\pm )*}\ad
$,
and spectrum $\sigma\ad^c$ is absolutely continuous.}
%%%%%%%%%%%%%%%%%%%%%%%%%%%%%%%%%%%%%%%%%%%%%%%%%%%%%%%%%%%%%%%%%%%%%%%%%%
\section{\hspace*{-1em}. INNER PRODUCT MAKING NEW HAMILTONIANS SELF-ADJOINT}
\label{InnerProduct}
%%%%%%%%%%%%%%%%%%%%%%%%%%%%%%%%%%%%%%%%%%%%%%%%%%%%%%%%%%%%%%%%%%%%%%%%%%
We introduce now a new inner product $[\, .\, ,\, .\,]\ad$  in
$
             {\cH} \ad  ,
$
%%%%
$
            [f\ad,g\ad]\ad= \langle  X\ad  f\ad,g\ad   \rangle,
$
$f\ad,g\ad\in\cH\ad,$
with $X\ad$ defined as in Theorem~\ref{ThInvariant},
$X\ad=I\ad +Q\abd Q\abd^{*}$,
$\an=1,2.$
  The operator $X\ad  $ is  positive  definite, $X\ad\geq I\ad$.
This means that $[\, .\, ,\, .\,]\ad$ satisfies all the  axioms  of
inner product.

\begin{theorem}\label{ThInnerProduct}\hspace*{-0.5em}{\sc :}
The operator $H\ad,  $ $\an=1,2,$  is self-adjoint on $\cD(A\ad)$
with respect to the inner product $[\, .\, ,\, .\,]\ad$.
\end{theorem}

\noindent{\sc Proof.} It follows from Theorem~\ref{ThInvariant}
that operator ${\bH}'$ is
self-adjoint in $\cH=\cH_1 \oplus\cH_2 $ with respect to the inner product
$[\, .\, ,\, .\,]$, $[f,g]=[Xf,g]$ with $X=\diag\{ X_1, X_2 \}$. Did, since
$\cQ^{-1}=\cQ^{*}X^{-1}=X^{-1}\cQ^{*}$, we have for
$f,g\in\cD({\bH}')=\cD({\bH})=\cD(A_1)\oplus\cD(A_2)$:
$$
    [{\bH}'f,g]=\langle    X\cQ^{-1} {\bH} \cQ f,g  \rangle =
     \langle  X\cdot X^{-1} \cQ^{*}{\bH}\cQ f, g \rangle   =
$$
$$
   = \langle f, \cQ^{*} {\bH}\cQ g  \rangle  =
     \langle f, X\cdot X^{-1} \cQ^{*} {\bH} \cQ g \rangle   =
     [f,{\bH}'g].
$$
Here, we used the fact that in the case of~(\ref{resolvest}),
$\cQ f \in \cD(A_1)\oplus\cD(A_2)$
if
$ f \in \cD(A_1)\oplus\cD(A_2).$

Taking elements $f,$ $g$ in the equality $[{\bH}'f,g]=[f,{\bH}'g]$
in the form $f=\{  f_1, 0  \}$, $g=\{ g_1, 0  \}$
or $f=\{ 0, f_2  \}$, $g=\{  0, g_2 \}$
with one of the components equal to zero and $f\ad,g\ad\in\cD(A\ad)$,
$\an=1,2$, one comes to the statement of Theorem.
\medskip

\noindent{\sc Remark.} This Theorem may be proved also in another way
making use of the equality
\be
\label{QQ}
               I\ad + Q\abd Q\abd^{*}  =
           \Sum_{j:\,\, U_j\in\cU \ad^d}
    \tpsi\ad\ju \langle\,\cdot\, ,\tpsi\ad\ju \rangle +
          \tPsipm\ad  \tPsipm {}^{*}\ad
\ee
which is valid for both signs ``$+$'' and ``$-$''.
In this case, a self-adjointness of $H\ad$ with respect to
$[\,\cdot\, , \,\cdot\, ]\ad$  follows
from the fact that it's spectrum is real and also from
relations
$
      H^{*}\ad  \tilde{\Psi }^{(\pm )}\ad
    \tilde{\Psi }^{(\pm )*}\ad  =
$
$
      \tilde{\Psi }^{(\pm )}\ad  A^{(0)}\ad
          \tilde{\Psi }^{(\pm )*}\ad  =
$
$
               \tilde{\Psi }^{(\pm )}\ad
         \tilde{\Psi }^{(\pm )*}\ad  H\ad
$.
The equality~(\ref{QQ}) itself is proved by calculating it's right part
in the same way as it was done when the completeness relations
(\ref{completeness}) were established
(see proof of Theorem~\ref{ThExpansion}).
%%%%%%%%%%%%%%%%%%%%%%%%%%%%%%%%%%%%%%%%%%%%%%%%%%%%%%%%%%%%%%%%%%%%%%%%%%
\section{\hspace*{-1em}. SCATTERING PROBLEM}\label{ScatteringProblem}
%%%%%%%%%%%%%%%%%%%%%%%%%%%%%%%%%%%%%%%%%%%%%%%%%%%%%%%%%%%%%%%%%%%%%%%%%%
We establish now that operators
$
\Psi\ad^{(+)}
$
and
$
\Psi\ad^{(-)}
$
play the same important role describing a time asymptotics of solutions
of the Schr\"{o}dinger equation
\be
\label{Schrod}
i\Frac{d}{dt}f\ad(t)=H\ad f\ad(t)
\ee
 as in the usual self-adjoint case\ct{LadyzhFaddeev},\ct{Faddeev64}.

%%%%%%%%%%%%%%%%%%%%%%%%%%%%%%%%%%%%%%%%%%%%%%%%%%%%%%%%%%%%%%%%%%%%%%%
\begin{theorem}\label{ThWave}\hspace*{-0.5em}{\sc :}
%%%%%%%%%%%%%%%%%%%%%%%%%%%%%%%%%%%%%%%%%%%%%%%%%%%%%%%%%%%%%%%%%%%%%%%%
Operator $U\ad(t)=\exp(iH\ad t)\exp(-iA\ad^{(0)} t)$
converges strongly if $t\rightarrow\mp\infty$,
with respect to the norm
$\parallel\cdot\parallel\ad^X$
corresponding to the inner product $[\,\cdot\, ,\,\cdot \,]\ad$
in ${\cH} \ad$. The limit is equal to
$\,\,\, s\!-\!\Lim{t\rightarrow\mp\infty} U\ad(t)=\Psi\ad^{(\pm)}.$
\end{theorem}

Since the norms $\| \,\cdot\, \|^{X}\ad $ and $\| \,\cdot\,\| $ in $\cH\ad$
are equivalent,
$\|f\|\leq\|f\|^{X}\ad \leq (1+\|Q\abd\|\cdot\|Q\bad\|)^{1/2}\|f\|$,
the same statement takes place also with respect to the initial norm
$\| \,\cdot\,\| $.

%%%%%%%%%%%%%%%%%%%%%%%%%%%%%%%%%%%%%%%%%%%%%%%%%%%%%%%%%%%%%%%%%%%%%%%%%
\begin{theorem}\label{ThScattering}\hspace*{-0.5em}{\sc :}
%%%%%%%%%%%%%%%%%%%%%%%%%%%%%%%%%%%%%%%%%%%%%%%%%%%%%%%%%%%%%%%%%%%%%%%%%
For any element $f\ad^{(-)}\in\cH\ad^c$
one can find such unique element  $f\ad^{(0)}$
that solution
$f\ad(t)=\exp(-iH\ad t)f\ad^{(0)}$
of Eq.~(\ref{Schrod})
satisfies the asymptotic condition
$$
\Lim{t\rightarrow-\infty}\parallel{f\ad(t)-\exp(-iA\ad^{(0)}t)
f\ad^{(-)}}\parallel\ad^X=0.
$$
There exists the unique element $f\ad^{(+)}\in\cH\ad^c$ such that
$$
\Lim{t\rightarrow+\infty}\parallel{f\ad(t)-\exp(-iA\ad^{(0)}t)
f\ad^{(+)}}\parallel\ad^X=0.
$$
Elements $f\ad^{(-)}$ and $f\ad^{(+)}$ are connected by the relation
$f\ad^{(+)}=S^{(\an)}f\ad^{(-)}$
with
\newline
$
     S^{(\an)}=
$
$
      \Psi\ad^{(-)-1}\Psi\ad^{(+)}=
$
$
       \tilde{\Psi}\ad^{(-)*}\Psi\ad^{(+)}=
$
$
         \Psi\ad^{(-)*} X\ad \Psi\ad^{(+)}.
$
\end{theorem}

We do not give here proofs of the Theorems~\ref{ThWave}
and~\ref{ThScattering} because they are exactly
the same as in the case of one-particle Schr\"{o}dinger operator in
Ref.~\ct{Faddeev63}

Theorem~\ref{ThScattering}  gives the non-stationary formulation of the scattering problem
for a system described by Hamiltonian $H\ad$.
Moreover $S^{(\an)}$ is a scattering   operator for this system.

%%%%%%%%%%%%%%%%%%%%%%%%%%%%%%%%%%%%%%%%%%%%%%%%%%%%%%%%%%%%%%%%%%%%%%%%%
\begin{theorem}\label{ThScChannel}\hspace*{-0.5em}{\sc :}
%%%%%%%%%%%%%%%%%%%%%%%%%%%%%%%%%%%%%%%%%%%%%%%%%%%%%%%%%%%%%%%%%%%%%%%%%
Scattering operator $S^{(\an)}$ coincides with the component
$s\aad$ of the scattering operator $S$, $S=U^{(-)*}U^{(+)}$,
for a system described by the two--channel Hamiltonian ${\bH}$.
\end{theorem}

\noindent{\sc Proof.} Let us show that operator $S^{(\an)}$
has the kernel $s\aad(\lambda,\lambda')$ given by Eq.~(\ref{scattering}).
To do this, remember that $\tPsi\ad^{(-)}=\Psi\ad^{(-)}-Q\abd u\bad^{(-)}$
(see Theorem~\ref{ThExpansion}). Therefore,
$$
S^{(\an)}=(\Psi\ad^{(-)*}-u\bad^{(-)*} Q\abd^{*})\Psi\ad^{(+)}=
\Psi\ad^{(-)*} \Psi\ad^{(+)} + u\bad^{(-)*} Q\bad \Psi\ad^{(+)} =
u\aad^{(-)*} u\aad^{(+)} +  u\bad^{(-)*} u\bad^{(+)} .
$$
Here, we have used the properties
$\Psipm\ad = u\aad^{(\pm)} $,
$Q\abd^{*}=-Q\abd$ and
$Q\bad \Psi\ad^{(+)} =u\bad^{(+)}$ established above.
Since
$$
            u\aad^{(-)*} u\aad^{(+)} +  u\bad^{(-)*} u\bad^{(+)} =
          \left(  U^{(-)*}U^{(+)}  \right) \aad = s\aad,
$$
we come to the statement of Theorem. The proof of
Theorem~\ref{ThScattering} is completed.
\medskip

A kernel of the scattering operator $S^{(\an)}$ may be presented
also in a usual way~(\ref{scattering}) in terms of the $t$-matrix
$t\ad(z)=W\ad-W\ad(H\ad-z)^{-1}W\ad$, taken on the energy--shell.
Note that $t\ad(z)$ differs from $T\aad(z)$ introduced in Sec.~\ref{Initial}.
Did, easy calculations show that
\be
\label{Tmatrix}
t\ad(z)=B\abd [I\bd +Q\bad (A\ad -z)^{-1} B\abd]^{-1} Q\bad .
\ee
Using the basic equation~(\ref{QbasicSym}) one can rewrite~(\ref{Tmatrix})
in the form
$$
    t\ad(z)=T\aad(z) + \tilde{t}\ad(z)
$$
where
$$
\tilde{t}\ad(z)= B\abd [A\bd -
B\bad(A\ad -z)^{-1}B\abd]^{-1}Q\abd (A\ad -z)\not\equiv 0.
$$
However the additional term $\tilde{t}\ad(z)$
is evidently disappearing on the energy-shell due to presence of the
difference $A\ad -z$ as an end factor. Actually, in the
diagonal representation~(\ref{Neumann}),(\ref{multi}),
$A\ad -z$ acts as the factor $\lambda-z$ vanishing at $z=\lambda+i0$.
Therefore, kernels of $t$-matrices $t\ad$ and $T\aad$ coincide
on the energy surface.

Note also that in our case
$\sigma_1^c\bigcap\sigma_2^c=\emptyset$. Hence we have
$s_{\bn\an}=0$ and $S^{(\an)}=s_{\an\an}$ is unitary.
%%%%%%%%%%%%%%%%%%%%%%%%%%%%%%%%%%%%%%%%%%%%%%%%%%%%%%%%%%%%%%%%%%%%%%%%%%
\section{\hspace*{-1em}. ON USE OF
THE TWO--BODY ENERGY--DEPENDENT POTENTIALS IN FEW--BODY PROBLEMS}
\label{2bodyPotNbody}
%%%%%%%%%%%%%%%%%%%%%%%%%%%%%%%%%%%%%%%%%%%%%%%%%%%%%%%%%%%%%%%%%%%%%%%%%
There is a rather conceptual question (see for instance
Refs.~\ct{McKellarMcKay}, \ct{Schmid}) concerning a use of the
two--body energy--dependent potentials in few--body non--relativistic
scattering problems. Evidently this question is strongly related
to the subject of the paper and we will discuss here three approaches
seemed to be reasonable when one tries to embed energy--dependent
potentials in few--body equations.

A customary way to embed such
potentials into the center-of-mass frame
N--body Schr\"odinger equation consists in the following. Namely, one
replaces
(see papers~\ct{NarodetskyPalanga},~\ct{OrlowskiKim},
\ct{YaF88},~\ct{KMMMP}
and Refs. therein)
the pair energy $z_{ij}$,
argument of the potential $V_{ij}(z_{ij})$, $i\neq j,$
describing interaction in  two-body subsystem
$\{ i,j \}$  ($i,j$ stand for numbers of particles, $i,j=1,2,...,$N)
with the  difference $Z - T'_{ij}$
between total energy $Z$ of system and
the kinetic energy operator $T'_{ij}$
of particles, supplementary  to the  subsystem $\{ i,j \}$.
For the resolvent-like energy dependent potentials~(\ref{epot})
this replacement is quite correct from mathematical point of view since
one can reconstruct underlying multichannel (for instance, four--channel
if N=3) self-adjoint Hamiltonian~\ct{YaF88},~\ct{KMMMP}. Reducing the
spectral problem for this Hamiltonian to the external channel only
one gets the N--body Schr\"{o}dinger equation exactly with the pair
potentials $V_{ij}(Z-T'_{ij})$. Thus one can guarantee that spectrum of this
equation is real and the scattering problem for the N--body system can be
based.

However the replacements $z_{ij}\rightarrow Z-T'_{ij}$ meet serious
conceptual objections formulated in concentrated form by
E.W.Schmid~\ct{Schmid}. Did,  it follows from the energy
conservation law that to obtain a share of total energy belonging to
subsystem $\{ i,j \}$, one has to subtract from $Z$
not only $T'_{ij}$ but
also a potential energy  of interaction between particles
$i,j$ and the rest particles of the system.
This idea shows really a first way for
the correct (in the context of Ref.~\ct{Schmid})
embedding two-body potentials  into N-body
equations: one has to redefine pair potentials as solutions
${\bf V}_{ij}$ of the following system of equations:
\be
\label{Nsystem}
  {\bf V}_{ij} = V_{ij} (Z - T'_{ij}
  - \Sum_{   \{i',j'\}   \neq  \{i,j\}   } {\bf V}_{i'j'}   )
\ee
where $\{i,j\}$ runs all the pair subsystems.
So, the usual embbedings $ V_{ij}(z_{ij})\rightarrow V_{ij}(Z-T'_{ij})$
may be considered only as a zero
approximation to solutions
${\bf V}_{ ij }(Z)$ of the system~(\ref{Nsystem}).
Unfortunately, this system may be treated
relatively easy only in the case of
linear dependence of the potentials $V_{ ij }(z_{ ij })$
on the (pair) energies $z_{ ij }$.
One can show in this case that operator--value functions
$V_{ij}(Y)$ of the operator variable $Y$,
$Y:\, L_2(\bR^{3(N-1)})\rightarrow L_2(\bR^{3(N-1)})$ may be defined in
such a way that solutions of Eqs.~(\ref{Nsystem})
generate only real spectrum for the N--body Schr\"{o}dinger
equation.

In the case of the resolvent-like energy dependence~(\ref{epot}) of pair
interactions one meets serious obstacles in solving
the system~(\ref{Nsystem}) connected with a strong non-linearity of it's
equations. Also, no underlying self--adjoint Hamiltonian is still
found.

Another way to deal with the two-body energy-dependent potentials
in few--body problems is to replace them with energy--independent ones.
In fact, in the present work we realized namely this idea which was
pronounced by B.H.J.McKellar and C.M.McKay~\ct{McKellarMcKay}. Did,
let us denote now a ``share'' of the total
energy of the N-body system belonging to the pair
subsystem $\{ i,j \}$, by $h_{ij}$.
Then this $h_{ij}$ has to satisfy the operator equation  following from
the  energy conservation law, too,
\be
\label{henind0}
   h_{ij}=h_{ij}^{(0)} +v_{ij} + V_{ij} (h_{ij}),
\ee
where $h_{ij}^{(0)}$ stands for the kinetic energy operator of the
pair $\{i,j\}$ and $v_{ij}$,
for an energy-independent part of the pair interaction.
Remember that the equation~(\ref{henind0}) in notation~(\ref{basic}) was
a main subject of the present work.
If solutions $h_{ij}$ of equations~(\ref{henind0}) be known,
one could substitute the (energy-independent) operators
$ W_{ij}\equiv V_{ij} (h^{ij} ) $ in the N-body
Hamiltonian treating them
in conventional way as additional energy-independent potentials.
It should be noted however that the potentials $W_{ij}$ are not totally
equivalent to the potentials $V_{ij}(z)$ given by~(\ref{epot}) since the
Hamiltonian $h_{ij}$ being solution of~(\ref{henind0}),
reproduces only a part of spectrum of the Schr\"{o}dinger equation
\be
\label{Sch3}
(h_{ij}^{(0)} + v_{ij} +V_{ij}(z))\psi=z\psi
\ee
(see Sec.~\ref{Eigenfunctions}).
Forbidden eigenstates correspond normally to the spectrum
generated by respective internal Hamiltonian~\ct{MotJMPh91}.
There is also another question: is the spectrum of the N--body Hamiltonian
real if potentials $W_{ij}$ are substituted in? Thing is that Hamiltonian
$h_{ij}$ becomes self-adjoint only with respect to a new inner product
in $L_2(\bR^3)$ (see Sec.~\ref{InnerProduct}). One can overcome this difficulty replacing
$h_{ij}$ with similar Hamiltonian
$h'_{ij}=X_{ij}^{1/2} h_{ij} X_{ij}^{-1/2}$ where $X_{ij}$ is analog for
$h_{ij}$ of the operators $X\ad$ introduced in Sec.~\ref{BasicEquation}. Writing $h'_{ij}$
in the form $h'_{ij} = h_{ij}^{(0)} +V'_{ij}$ one gets a new pair potential
$V'_{ij}$ which is already self--adjoint with respect to the standard
inner product in $L_2(\bR^3)$. Thus, one may use then the potentials
$V'_{ij}$ being sure that the  N--body Hamiltonian constructed is Hermitic.
Emphasize that potential $V'_{ij}$ gives the same two--body spectrum and
phase shifts as the potential $v_{ij}+W_{ij}$ because $h'_{ij}$ is obtained
from $h_{ij}$ by similarity transform. It follows from
Theorems~\ref{ThScattering} and~\ref{ThScChannel} that
the phase shifts given by $V'_{ij}$ coincide also with those for
Eq.~(\ref{Sch3}). Therefore, the operator $V'_{ij}$ turns out one
of the phase--equivalent potentials for the two--body subsystem concerned.

So, we have discussed three different approaches to embedding the two-body
energy--dependent potentials in few--body problems. Certainly, the
approaches based on solving the non-linear equations~(\ref{Nsystem})
and~(\ref{henind0}) do not seem to be too attractive from the computational
point of view. However, in the cases when the internal Hamiltonians of
pair subsystems have a finite discrete spectrum only and the coupling of
channels is relatively small (see Corollary~\ref{CBestimate}
to Theorem~\ref{ThSolvability}),  the approach
based on solving Eqs.~(\ref{henind0}) could be quite realized numerically.

%%%%%%%%%%%%%%%%%%%%%%%%%%%%%%%%%%%%%%%%%%%%%%%%%%%%%%%%%%%%%%%%%%%%%%%%%%
\medskip
	
\noindent{\bf ACKNOWLEDGEMENTS}
\medskip

%%%%%%%%%%%%%%%%%%%%%%%%%%%%%%%%%%%%%%%%%%%%%%%%%%%%%%%%%%%%%%%%%%%%%%%%%
Author is thankful to Prof. B.S.Pavlov and Dr.~K.A.Makarov
for support and interest to this work.
The author is much indebted to Prof. R.A.Minlos and  participants
of his seminar in Moscow University for discussion of presented results. 
Particularly, the author is grateful to Dr.~S.A.Stepin for the literary 
indications.
\small
\baselineskip12pt
%%%%%%%%%%%%%%%%%%%%%%%%%%%%%%%%%%%%%%%%%%%%%%%%%%%%%%%%%%%%%%%%%%%%%%%%%%%%

\end{document}